\documentclass[aps,pre,preprint,superscriptaddress,amsmath,amssymb,onecolumn,11pt]{revtex4}
\usepackage{graphicx}
\usepackage{amsmath}
\usepackage{hyperref}
\usepackage{color}
\usepackage{bigints}
\usepackage{mathrsfs}
\usepackage{algorithm}
\usepackage{algpseudocode}
\definecolor{darkblue}{RGB}{8,81,156}
\hypersetup{colorlinks,breaklinks,
            linkcolor=darkblue,urlcolor=darkblue,
           anchorcolor=darkblue,citecolor=darkblue}

\date{\today}

\allowdisplaybreaks

    \definecolor{dark-purple}{RGB}{118,42,131}
    \definecolor{dark-green}{RGB}{27,120,55}
    \definecolor{light-purple}{RGB}{231,212,232}
    \definecolor{LIGHT-PURPLE}{RGB}{194,165,207}
    \definecolor{light-green}{RGB}{168,216,183}
    \definecolor{gray}{RGB}{186,186,186}
    \definecolor{super-dark-green}{RGB}{0,69,41}
    \definecolor{super-dark-purple}{RGB}{63,0,125}
    \definecolor{super-dark-blue}{RGB}{8,48,107}
    \definecolor{super-dark-red}{RGB}{165,0,38}
    
    \definecolor{super-dark-purple}{RGB}{64,0,75}
    \definecolor{super-dark-green}{RGB}{0,68,27}

\usepackage{tabularx,booktabs} 

\usepackage{array}
\usepackage{longtable}
\newcolumntype{L}[1]{>{\raggedright\let\newline\\\arraybackslash\hspace{0pt}}p{#1}}
\newcolumntype{C}[1]{>{\centering\let\newline\\\arraybackslash\hspace{0pt}}m{#1}}
\newcolumntype{R}[1]{>{\raggedleft\let\newline\\\arraybackslash\hspace{0pt}}m{#1}}

\begin{document}

\title{Forward-Flux Sampling with Jumpy Order Parameters}

\author{Amir Haji-Akbari}
\email{amir.hajiakbaribalou@yale.edu}
\affiliation{Department of Chemical and Environmental Engineering, Yale University, New Haven, CT  06520}

%\author{Pablo G. Debenedetti}
%\email{pdebene@exchange.princeton.edu}
%\affiliation{Department of Chemical and Biological Engineering, Princeton University, Princeton, NJ 08540}
%
\date{\today}

\begin{abstract}
Forward-flux sampling (FFS) is a path sampling technique that has gained increased popularity in recent years, and has been used to compute rates of rare event phenomena such as crystallization, condensation, hydrophobic evaporation, DNA hybridization and protein folding. The popularity of FFS is not only due to its ease of implementation, but also because it is not very sensitive to the particular choice of an order parameter. The order parameter utilized in conventional FFS, however, still needs to satisfy a stringent smoothness criterion in order to assure sequential crossing of FFS milestones. This condition is usually violated for order parameters utilized for describing aggregation phenomena such as crystallization. Here, we present a generalized FFS algorithm for which this smoothness criterion is no longer necessary, and apply it to compute homogeneous crystal nucleation rates in several systems. Our numerical tests reveal that conventional FFS can sometimes underestimate the nucleation rate by several orders of magnitude.
\end{abstract}

\maketitle

\section{Introduction\label{section:intro}}

Rare events are ubiquitous in nature, and their occurrence is predicated upon the emergence of highly improbable fluctuations in the system. The separation of timescales between the time needed for the emergence of a favorable fluctuation, and  the actual duration of the ensuing rare event makes it impractical-- if not impossible-- to capture its kinetics and microscopic mechanism using conventional sampling techniques such as molecular dynamics (MD)~\cite{AlderMDJCP1959} or Monte Carlo (MC)~\cite{Metropolis1953}. Instead, advanced path sampling techniques are necessary to obtain a statistically representative ensemble of reactive trajectories~\cite{ChandlerTPS2002, LaioPNAS2002,  vanErpJChemPhys2003, MaraglianoJChemPhys2006, FrenkelFFS_JCP2006, AdamsPRL2008, AdamsPhysRevE2009, AdamsEurophysLett2009, AdamsJCP2010, AdamsJCP2_2010, BeckerJChemPhys2012, TiwaryPRL2013, SantisoJChemPhys2015,  GotchyJCTC2015}. One such algorithm that has gained increased popularity in recent years is forward-flux sampling (FFS)~\cite{AllenFrenkel2006} in which the cumulative flux of reactive trajectories is computed along an \emph{order parameter} $\lambda:\mathscr{Q}\rightarrow\mathbb{R}$ that quantifies the progress of the transition from $A:=\left\{x\in\mathscr{Q}: \lambda(x)<\lambda_A\right\}$ to $B:=\left\{x\in\mathscr{Q}: \lambda(x)\ge\lambda_B\right\}$. Here, $\mathscr{Q}$ is the configuration space of the underlying system, and $A$ and $B$ are two of its local free energy minima. In recent years, FFS  has been used for studying a wide range of rare-event-driven phenomena such as evaporation~\cite{SumitPNAS2012, MeadleyJCP2012, AltabetPNAS2017}, coalescence~\cite{RekvigJCP2007}, wetting~\cite{SavoyLangmuir2012}, magnetic switching~\cite{VoglerPRB2013}, protein folding~\cite{BorreroJCP2006}, DNA hybridization~\cite{OuldridgeNucleicAcidRes2013, HinckleyJChemPhys2013}, phase separation in active systems~\cite{RichardSoftMatter2016}, protein aggregation~\cite{SmitJPhysChemB2016} and crystal nucleation~\cite{SanzPRL2007, LiNatMater2009, FillionJCP2010, GalliPCCP2011, GalliNatComm2013, HajiAkbariFilmMolinero2014,  ThaparPRL2014,  HajiAkbariPNAS2015, CabrioluPRE2015, GianettiPCCP2016, SossoJPhysChemLett2016, BiJPhysChemC2016, BiJChemPhys2016, HajiAkbariPNAS2017}.  

A major ambiguity in applying path sampling techniques arises from the fact that most rare events can be satisfactorily described by more than one order parameter. FFS is particularly insensitive  to this degeneracy, and a subpar order parameter only compromises its efficiency and not its accuracy~\cite{BorreroJCP2007}. 
Despite this flexibility, $\lambda(\cdot)$ still needs to satisfy a stringent smoothness criterion i.e.,~$\lambda(t)$ should not undergo big fluctuations along a discrete-time trajectory. A sufficient-- but not necessary-- condition for smoothness is the uniform continuity of $\lambda$ in $\mathscr{Q}$, which assures that fluctuations in $\lambda(t)$ can be bounded, e.g.,~by choosing a sufficiently small time step.
We denote an order parameter that is not smooth as 'jumpy`. In other words, the value of a jumpy order parameter can undergo large changes even after a single MD time step or MC sweep. For a  real-valued $\lambda(\cdot)$, jumpiness usually involves the existence of discontinuities-- or possibly the lack of uniform continuity-- in $\lambda(x)$. For an integer-valued order parameter, jumpiness refers to the possibility that $\lambda(x)$ can change by more than $\pm1$ between successive MD time steps or MC sweeps.
 It is relatively straightforward to identify smooth order parameters for phenomena such as hydrophobic evaporation and protein folding. For aggregation phenomena, such as crystallization and phase separation, however, almost all existing order parameters violate  smoothness, and the accuracy of conventional FFS is therefore not guaranteed. The violation arises from the underlying physics of aggregation phenomena that involve the coalescence of subcritical nuclei of the new phase within the exiting metastable phase.
For instance, if the number of atoms/molecules within the largest nucleus of the new phase is defined as the order parameter, it will jump by the number of atoms/molecules within a smaller nucleus that coalesces to the largest nucleus in the system.  In this paper, we develop a generalized variant of FFS, which we call jumpy FFS (jFFS), for which this smoothness criterion is no longer necessary. We numerically compare the rates computed from jFFS and conventional FFS (cFFS) and conclude that the latter can systematically underestimate the nucleation rate, sometimes by several orders of magnitude. 

This paper is organized as follow. In Section~\ref{section:qual}, we provide a qualitative description of jFFS and how it is different from conventional FFS, while a mathematically rigorous derivation of jFFS is provided in Section~\ref{section:mathematical}. Section~\ref{section:methods} is dedicated to the technical details of the nucleation rate calculations, with the results presented in Section~\ref{section:results}. Finally, Section~\ref{section:summary} is reserved for summary and concluding remarks.

\section{Qualitative Description of jFFS}
\label{section:qual}
The whole premise of the conventional FFS algorithm is to carry out the transition from $A$ to $B$ in stages by placing $N$ milestones, $\lambda_A<\lambda_0<\lambda_1<\cdots<\lambda_{N}=\lambda_B$, between the two basins. The flux of trajectories that cross each milestone is then computed recursively as follows.  First, a sufficiently long trajectory is generated in $A$ using conventional unbiased techniques such as molecular dynamics or Monte Carlo, in order to compute the flux of trajectories that cross $\lambda_0$ after leaving $A$. In general, $\lambda_0$ is chosen to be sufficiently close to $A$ so that it is crossed fairly frequently by such a trajectory. Whenever a crossing occurs, the corresponding configuration is stored for future iterations. $N_c$, the number of such crossings, is then used to calculate $\Phi_0=N_c/t$, the flux of trajectories originating $A$ and crossing $\lambda_0$. Here, $t$ is the total length of MD (or MC) trajectory utilized for such analysis. In many applications, $\Phi_0$ is further normalized by the average volume (or area) of the corresponding system. The second stage of the FFS algorithm involves $N$ iterations aimed at computing the transition probabilities between successive milestones. During the $k$th such iteration ($0\le k\le N-1$), for instance, a large number of trial trajectories are initiated from the configurations stored at $\lambda_k$. For $k=0$, the iteration uses the configurations collected during the long MD (or MC) trajectory in the  basin. Each trial trajectory is initiated from a randomly chosen configuration, and is propagated until it hits either of the $\lambda(x)=\lambda_{k+1}$ and $\lambda(x)=\lambda_A$ absorbing interfaces.  In order to make trial trajectories initiated from the same configurations distinct, degrees of freedom orthogonal to what goes into calculating $\lambda$ need to be properly randomized prior to propagating the trajectory. A proper procedure for randomizing is similar to what is conducted for hybrid Monte Carlo (HMC)~\cite{DuanePhysLettB1987}, which is discussed in detail in Ref.~\cite{PalmerJChemPhys2018}. For MD trajectories, this usually involves randomizing momenta according to the Maxwell-Boltzmann distribution, while in MC, it is just sufficient to use a new set of random numbers for conducting trial moves. The transition probability $P(\lambda_{k+1}|\lambda_k)$ is then estimated as the fraction of trajectories starting at $\lambda_k$ that reach $\lambda_{k+1}$ prior to returning to $A$. The flux of trajectories that cross $\lambda_0$ after leaving $A$ and the individual transition probabilities are then lumped together to obtain an estimate of the nucleation rate $R=\Phi_0\prod_{k=0}^{N-1}P(\lambda_{k+1}|\lambda_k)$, which is the cumulative flux of trajectories reaching $\lambda_N$ after leaving $A$.

The ability of conventional FFS to accurately predict rate is predicated on the condition that individual milestones are crossed sequentially, i.e.,~a trajectory starting from $\lambda_{k-1}$ never crosses $\lambda_{k+1}$ before crossing $\lambda_k$ at some earlier time. This condition implies that a configuration obtained upon crossing $\lambda_k$: (i) will always be at (or very close to) $\lambda_k$, (ii) can only be obtained from \emph{exactly} $k$ FFS iterations, $\lambda_0\rightarrow\lambda_1$, $\lambda_1\rightarrow\lambda_2, \cdots, \lambda_{k-1}\rightarrow\lambda_k$. It is easy to observe that this sequential crossing condition will be readily satisfied for a smooth order parameter.
%, i.e.,~a $\lambda(\cdot)$ that does not undergo large changes between successive MD steps (or MC sweeps). 
For order parameters that are jumpy, however, milestones will not necessarily  be   crossed sequentially, and, as a result, a configuration that is obtained upon crossing $\lambda_k$ might not only be closer to $\lambda_{k+1}$, but might also have been obtained from \emph{less than} $k$ iterations. We will refer to the sequence of iterations resulting in configuration $x$ as its \emph{jump history}.

\begin{figure}
\centering
\includegraphics[width=.42\textwidth]{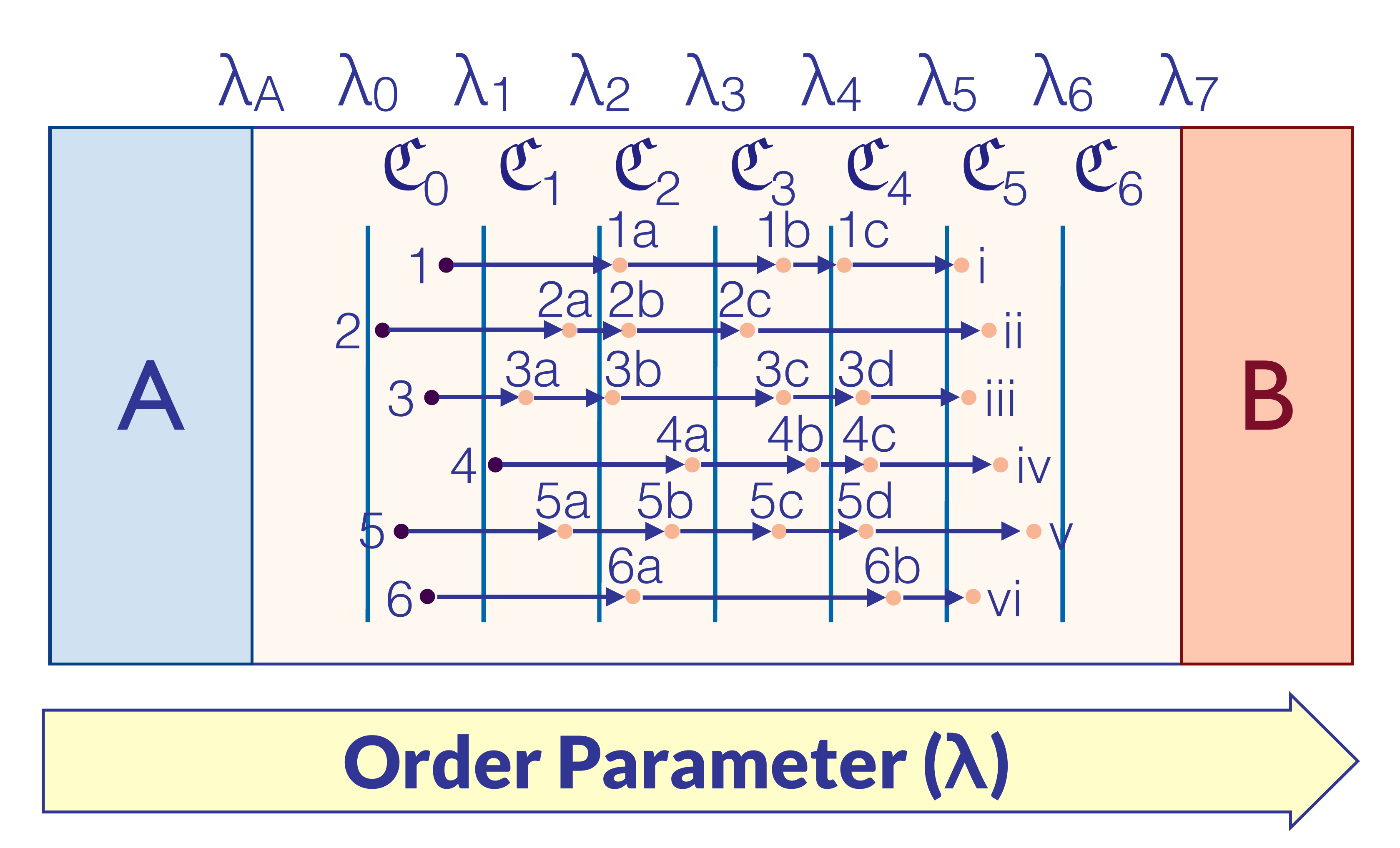}
\caption{\label{fig:jump-schematic} 
A schematic representation of what happens when a jumpy order parameter is utilized with conventional FFS. Purple configurations are obtained from crossings of $\lambda_0$ by a long trajectory in $A$, while each orange configuration is the endpoint of a trajectory initiated from the configuration connected to it with an arrow. For instance, 6b is the configuration corresponding to  crossing of $\lambda_3$ by a trajectory initiated from 6a. 
%A schematic representation of possible jump histories in FFS with a jumpy order parameter. 
}
\end{figure}

For instance, consider an FFS calculation with $N=7$ milestones between $\lambda_A$ and $\lambda_B$, conducted using a jumpy order parameter (Fig.~\ref{fig:jump-schematic}). The purple configurations, $1-6$, correspond to crossings of $\lambda_0$ by a trajectory originated in $A$. Usually, such a crossing will result in a configuration in the interval $\mathfrak{C}_0=[\lambda_0,\lambda_1)$. But the jumpiness of $\lambda(\cdot)$ makes it possible for such a trajectory to directly cross into $\mathfrak{C}_1=[\lambda_1,\lambda_2)$ without ever going through $\mathfrak{C}_0$, e.g.,~resulting in a configuration such as (4). Similarly, the FFS trajectories initiated from any configuration might completely skip some intermediate $\mathfrak{C}_k$'s. For instance, the sample trajectory initiated from (1) completely skips $\mathfrak{C}_1$, and results in 1a upon crossing $\lambda_1$. Among the six configurations in $\mathfrak{C}_5$, for instance, only  (iii) and (v) have been obtained from trajectories that have crossed $\mathfrak{C}_0,\mathfrak{C}_1,\cdots,\mathfrak{C}_4$ sequentially. Furthermore, even if a trajectory initiated from a given $\lambda_k$ does not skip $\mathfrak{C}_{k+1}$ upon crossing $\lambda_{k+1}$, it might still be closer to $\lambda_{k+2}$ than the target milestone $\lambda_{k+1}$. For instance, (v) is closer to $\lambda_6$ than $\lambda_5$. Conventional FFS is not equipped with rigorous recipes to handle such scenarios. What is commonly practiced though is that in computing  $P(\lambda_{k+1}|\lambda_k)$, trial trajectories are initiated from all the configurations in $\mathfrak{C}_k$ that are at (or close to) $\lambda_k$, irrespective of their jump history.  In computing $P(\lambda_6|\lambda_5)$ in Fig.~\ref{fig:jump-schematic}, for instance, only  (i), (iii) and (vi) are included in the list of starting configurations, despite having distinct jump histories, and (ii), (iv) and (v) are excluded because of their distance from $\lambda_5$. The actual transition probability $P(\lambda_{k+1}|\lambda_k)$ is then estimated as the fraction of trial trajectories that cross $\lambda_{k+1}$, irrespective of the $\mathfrak{C}_l(l>k)$ that they reach immediately after such a crossing.  These are all \emph{ad hoc} choices that cannot be rigorously justified, and, as will be shown here, can result in  considerable underestimation of the rate of the corresponding rare event.

\begin{figure}
\centering
\includegraphics[width=.5\textwidth]{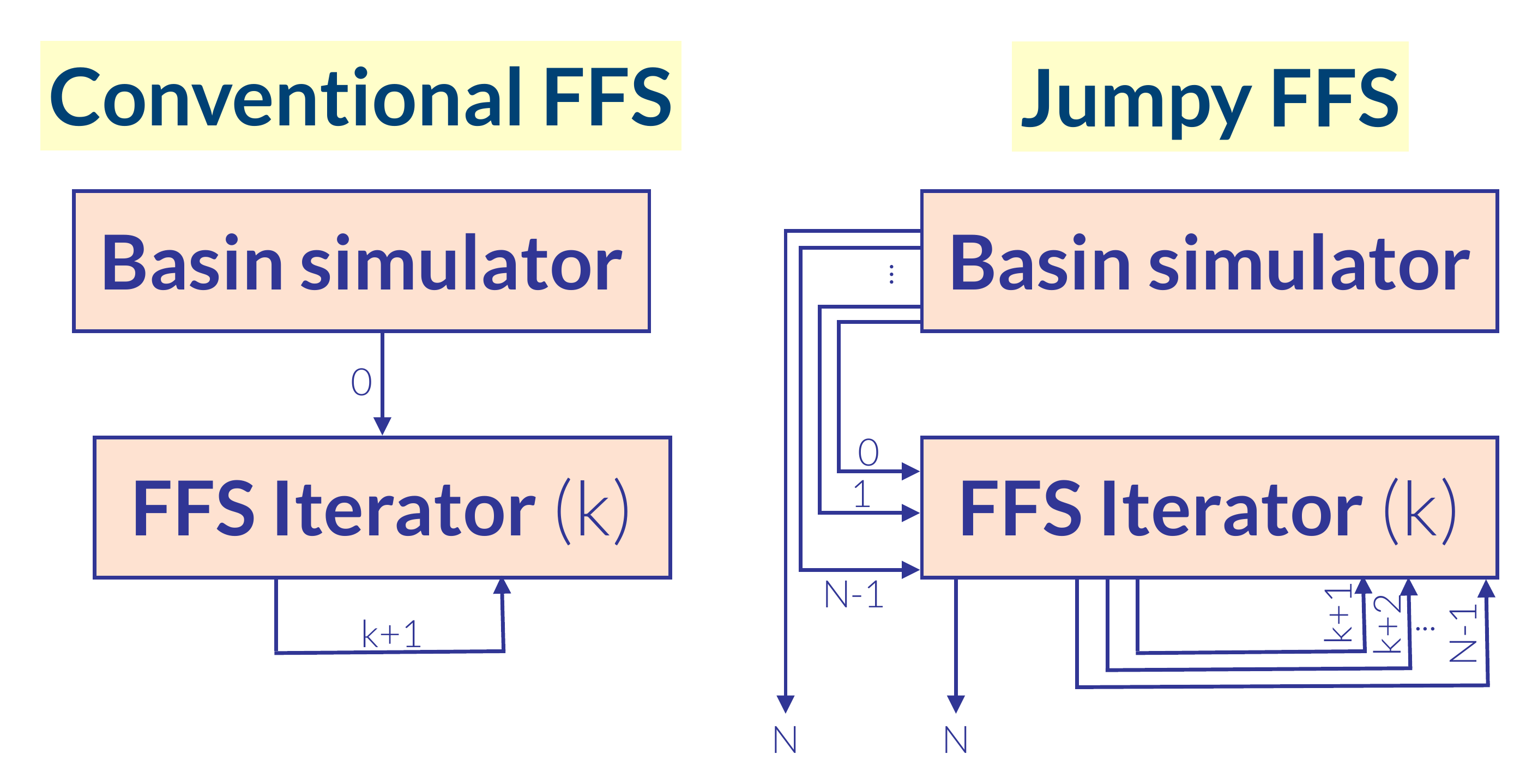}
\caption{\label{fig:ffs-scheme} A schematic representation of conventional FFS and jFFS, with the numbers appearing adjacent to each arrow referring to the landing index of the starting configurations sent to the FFS iterator. In conventional FFS, the configurations collected at each $\lambda_k$ are passed along to an FFS iteration aimed at crossing $\lambda_{k+1}$, while in jFFS, an iterations aimed at crossing $\lambda_k$ can generate configurations at any $\mathfrak{C}_l,~l\ge k$, which should then be passed along to an FFS iteration aimed at crossing the next milestone in line.
}
\end{figure}

The jFFS algorithm proposed here is a generalization of cFFS that properly accounts for such effects, and therefore accurately estimates the rate of a rare event described by a jumpy order parameter. A formal derivation of jFFS alongside its implementation details are provided in Section~\ref{section:mathematical}, but its main difference with cFFS is that each FFS iteration is initiated from a set of configurations that have the same jump history, and not those that have the same $\lambda$ value. If the configurations collected within a given $\mathfrak{C}_k$  have different jump histories, it will be necessary to conduct multiple FFS iterations from $\mathfrak{C}_k$, with each iteration initiated from configurations with the same jump history (Fig.~\ref{fig:ffs-scheme}). In Fig.~\ref{fig:jump-schematic}, for instance, the configurations in $\mathfrak{C}_5$ have five distinct jump histories, and therefore five distinct FFS iterations need to be initiated from $\mathfrak{C}_5$. Therefore, the notion of a transition probability between two "milestones`` is no longer meaningful in jFFS, as the configurations sharing a particular jump history might not all be necessarily close to the nominal starting milestone, and more importantly, the multiple iterations starting from the same $\mathfrak{C}_k$-- but from configurations with different jump histories-- might yield widely different transition probabilities (Fig.~\ref{fig:history-dependence}). According to combinatorics, a configuration in $B$ can, in principle, be obtained from reactive trajectories with $2^N$ distinct jump histories, and therefore the rate will no longer be a simple product of a flux and $N$ transition probabilities, but a sum of $2^N$ terms each corresponding to one of those $2^N$ distinct jump scenarios. In reality, however, not all $2^N$ jump scenarios are equally likely, since for most order parameters, large temporal fluctuations needed for multi-milestone jumps are extremely rare.  Furthermore, as we will explain in Section~\ref{section:mathematical}, FFS milestones can usually be chosen so that only one-- or at most a handful-- of those $2^N$ terms are nonzero. But as will be shown in Section~\ref{section:results}, even then, conventional FFS can underestimate the rate by several order of magnitudes, primarily due to not including the configurations that are far from the starting milestone in transition probability calculations.

\section{Formal Derivation of jFFS and Implementation Details}
\label{section:mathematical}

\subsection{Mathematical Derivation of jFFS}
\label{section:derive}
In order to rigorously describe the difference between  cFFS and jFFS, it is necessary to note that the goal of an FFS calculation is to sample $\mathcal{E}_A$, the ensemble of trajectories originating in $A$, and to estimate $\tau$, the average time that it takes for a trajectory in $\mathcal{E}_A$ to reach $B$. Here, a trajectory is a discrete sequence of configurations, $X\equiv(x_0,x_1,\cdots)\in\mathcal{E}_A$, propagated through a Markovian process, with the time-invariant transition probability $\pi(x_{n+1}|x_n)$. It is usually customary to report $\Phi_{A\rightarrow B}=1/\tau$, the average rate at which a trajectory in $\mathcal{E}_A$ reaches $B$. In certain applications, $\Phi_{A\rightarrow B}$ is also  normalized by the volume and/or the surface of the corresponding system. It is easy to observe that $\Phi_{A\rightarrow B}=\langle W_B\rangle_{\mathcal{E}_A}/\langle T_B\rangle_{\mathcal{E}_A}$, with $T_B[X]$ and $W_B[X]$ given by:
\begin{eqnarray}
T_B[X] &:=& \min_{q\ge L[X]} \{x_q\in A\cup B\} \\
L[X] &:=& \min_{q>0} \{x_q\not\in A\}\\
W_B [X] &:=& \theta_B (x_{T_B[X]})
\end{eqnarray}
Here, $T_B[X]\ge L[X]$ is the earliest time at which $X$, which has left $A$ at an earlier time $L[X]$, either reaches $B$ or returns to $A$, and $\theta_B(x)$ is an indicator function that is one if $x\in B$ and zero otherwise. For most rare events, $\langle W_B\rangle_{\mathcal{E}_A}$ is astronomically small, and cannot be estimated from direct sampling of $\mathcal{E}_A$. In FFS, $\langle W_B\rangle_{\mathcal{E}_A}$ is estimated by placing $N$ milestones, $\lambda_A=\lambda_{-1}<\lambda_0<\lambda_1<\cdots<\lambda_{N-1}<\lambda_N=\lambda_B$, between $A$ and $B$, and successively enumerating $T_i[X]$ and $U_{i,j}[X]$:
\begin{eqnarray}
T_i[X] &:=& \min_{q\ge L[X]} \{x_q\not\in\cup_{j=0}^{i}\mathfrak{C}_{j-1}\}\\
U_{i,j}[X] &:=& \left\{
\begin{array}{lll}
\theta_i(x_{T_i})\theta_j(x_{T_{i+1}}) &~~~~& i\ge0\\
\phi_0(x_L)\theta_j(x_{T_0})&~~~~& i=-1
\end{array}
\right.
\end{eqnarray}
with  $\mathfrak{C}_i=\{x\in\mathscr{Q}: \lambda_i\le\lambda(x)<\lambda_{i+1}\}$, $\theta_i(x)=\theta_{\mathfrak{C}_i}(x)$ and $\phi_i(x)=\sum_{j=0}^{i}\theta_{j-1}(x)$.
 In other words, $T_i[X]$ is the earliest time after $L[X]$ at which $X$ crosses $\lambda_i$ for the first time and $U_{i,j}[X]$  is a success indicator that specifies whether a trajectory that has already landed in $\mathfrak{C}_i$ as a result of crossing $\lambda_i$ at $T_i$ lands in $\mathfrak{C}_j$ at $T_{i+1}$. Note that if $x_{T_{i+1}}\in A$, $U_{i,j}=0$ for all $j>i$. Also if $U_{i,j}[X]=1$ for some $j>i+1$, $T_{i+1}[X]=\cdots=T_{j}[X]$ since under such a scenario, $\lambda_{i+2},\cdots,\lambda_j$ will also be crossed at the same time as $\lambda_{i+1}$.  The jump history of $x_{T_i}\not\in A$ can be formally defined as the ordered duplicate-free sequence $\mathfrak{h}(x_{T_i}):=\left[-1,s(x_{T_0}), s(x_{T_1}),\cdots,s(x_{T_i})\right]$, with $s(x)$ given by:
\begin{eqnarray}
s(x) &=& \left\{
\begin{array}{ll}
i~~~~~& x\in\mathfrak{C}_i\\
-1 & \lambda_A\le\lambda(x)<\lambda_0
\end{array}
\right.
\end{eqnarray}
In other words, $s(x)$ is the index of the region at which $x$ is located.
$W_B[X]$ can therefore be expressed as:
\begin{eqnarray}
W_B &=& U_{-1,N} +  \sum_{k=1}^{N}\sum_{-1<j_1<\cdots<j_k<N}U_{-1,j_1}U_{j_1,j_2}\cdots U_{j_k,N} \label{eq:W_B}
\end{eqnarray}
If $\lambda(\cdot)$ is smooth, $x_{T_i}\in\mathfrak{C}_i$ will either be at or very close to $\lambda_i$. Therefore, $U_{i,j}=0$ for $j>i+1$ and all but one term in Eq.~(\ref{eq:W_B}) will vanish. $\langle W_B\rangle_{\mathcal{E}_A}$ and $\Phi_{A\rightarrow B}^{\text{smooth}}$ will therefore be given by:
\begin{eqnarray}
\langle W_B\rangle_{\mathcal{E}_A}^{\text{smooth}} &=&  \langle U_{-1,0}\rangle_{\mathcal{E}_A}\prod_{q=0}^{N-1}\left\langle U_{q,q+1}|\left\{U_{r-1,r}=1\right\}_{r=0}^{q-1}\right\rangle_{\mathcal{E}_A}\notag\\
\Phi_{A\rightarrow B}^{\text{smooth}} &=& \Phi_{A\rightarrow\lambda_0}^{\text{smooth}}\prod_{q=0}^{N-1} P(\lambda_{q+1}|\lambda_q)
\end{eqnarray}
with $\Phi_{A\rightarrow\lambda_0}^{\text{smooth}}=\langle U_{-1,0}\rangle_{\mathcal{E}_A}/\langle T_B\rangle_{\mathcal{E}_A}$. For a rare event, however, $\langle T_B\rangle_{\mathcal{E}_A}$ is dominated by the trajectories returning to $A$ and therefore $\langle U_{-1,0}\rangle_{\mathcal{E}_A}/\langle T_B\rangle_{\mathcal{E}_A}\approx N_c/T_0$ with $N_c$ the number of crossings of $\lambda_0$ for a trajectory of length, $T_0$. 
Furthermore, since all crossings of a given $\lambda_q$ result in a configuration at or very close to $\lambda_q$,  $\langle U_{q,q+1}|U_{-1,0}=U_{0,1}=\cdots=U_{q-1,q}=1\rangle_{\mathcal{E}_A}$ reduces to $P(\lambda_{q+1}|\lambda_q)$. This is the familiar formalism of conventional FFS outlined in multiple earlier publications~\cite{AllenFrenkel2006, AllenFFSJCP2008, AllenFFSJCP2008} and explained in Section~\ref{section:qual}. 

\begin{figure}
\centering
\includegraphics[width=.5\textwidth]{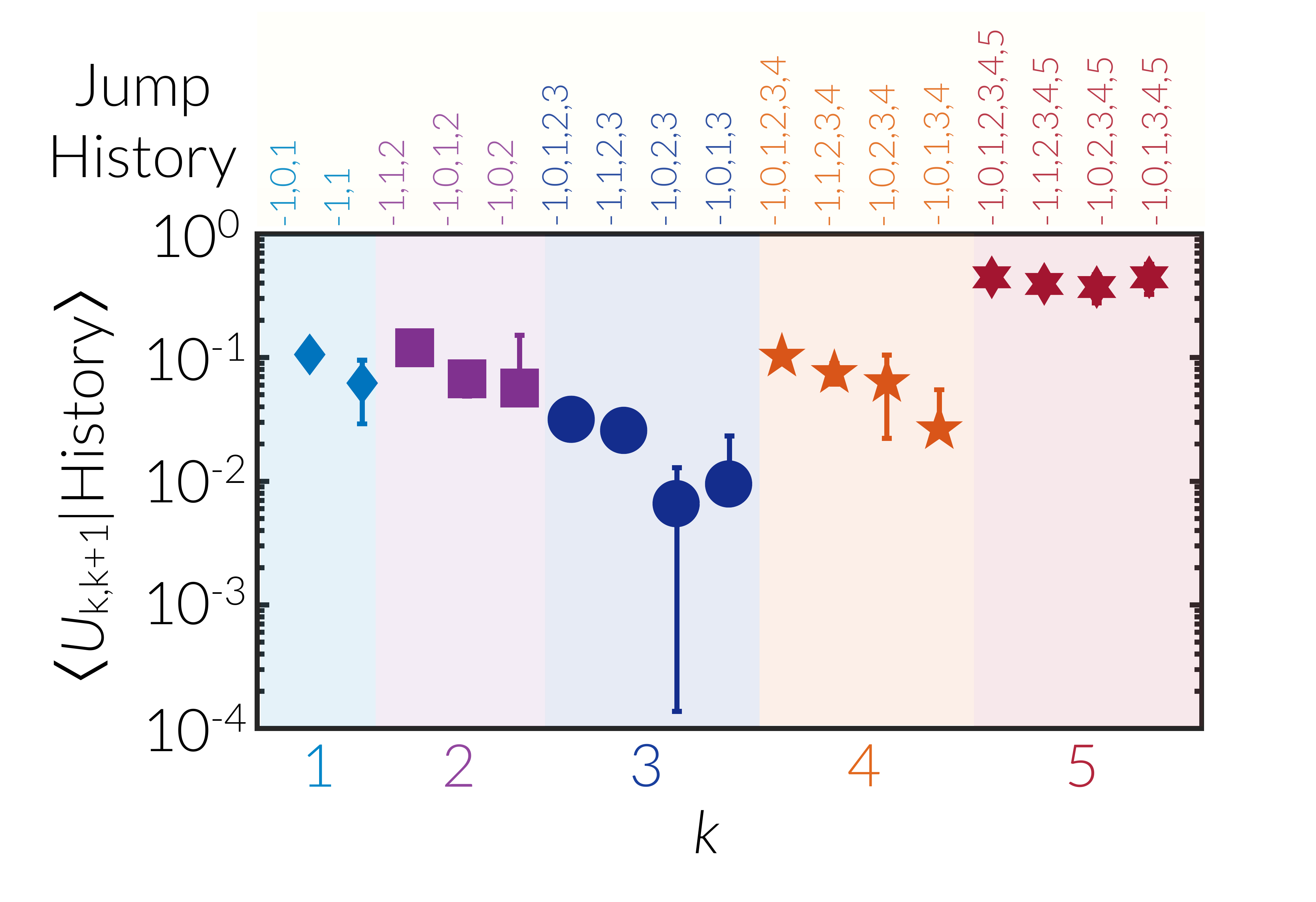}
\caption{\label{fig:history-dependence}
History dependence of transition probability for the homogeneous crystal nucleation rate calculation in the LJ system at $T^*=0.47$ and $p^*=0$. The entire calculation consisted of $N=7$ milestones with $(\lambda_{-1},\lambda_{0},\lambda_1,\cdots,\lambda_7)=(15,40,55,75,100,160,230,320,440)$. 
}
\end{figure}

For a jumpy $\lambda(\cdot)$, none of these assertions are necessarily true, and $U_{i,j}$ can be nonzero for any $j>i$. Therefore, each of the $2^N$ terms in Eq.~(\ref{eq:W_B}) can be nonzero, and can contribute to $\langle W_B\rangle_{\mathcal{E}_A}$. In general, $\left\langle \prod_{q=0}^{k}U_{j_q,j_{q+1}}\right\rangle_{\mathcal{E}_A}$ will be given by:
\begin{eqnarray}
\left\langle \prod_{q=0}^{k}U_{j_q,j_{q+1}}\right\rangle_{\mathcal{E}_A}&=& {\langle U_{-1,j_1}\rangle_{\mathcal{E}_A}}\prod_{q=1}^{k}\left\langle U_{j_q,j_{q+1}}|\left\{U_{j_{r-1},j_r}=1\right\}_{r=1}^{q-1}\right\rangle_{\mathcal{E}_A}\notag
\end{eqnarray}
Here, $\langle U_{j_q,j_{q+1}}|U_{-1,j_1}=\cdots=U_{j_{q-1},j_q}=1\rangle_{\mathcal{E}_A}$ is the probability that a trajectory propagated from a configuration $x\in\mathfrak{C}_{j_q}$ with jump history $\mathfrak{h}(x)=[-1,j_1,j_2,\cdots,j_q]$ ends up in $\mathfrak{C}_{j_{q+1}}$ at $T_{j_q+1}$. As outlined in Section~\ref{section:qual}, another important consequence of jumpiness is that $\lambda(x_{T_i})$ might be closer to $\lambda_{s(x_{T_i})+1}$ than $\lambda_{s(x_{T_i})}$. Therefore, unlike conventional FFS in which transition probabilities are computed from the configurations  at (or close to) $\lambda_i$, in jFFS all configurations in $\mathfrak{C}_i$ should be considered, even if they are closer to $\lambda_{i+1}$ than $\lambda_i$. Finally, transition probabilities will, in general, depend on the jump history of the starting configurations. In other words, $\langle U_{i,k}U_{k,l}\rangle/\langle U_{i,k}\rangle\neq\langle U_{j,k}U_{k,l}\rangle/\langle U_{j,k}\rangle$. We provide an analytical argument for this history dependence in Appendix~\ref{sec:Uij_derive}, and confirm it numerically through our jFFS calculations of the homogeneous crystal nucleation rate in the Lennard-Jones system (Fig.~\ref{fig:history-dependence}).

\subsection{Implementation Details}

We now describe the numerical procedure for estimating the expected values of terms in Eq.~(\ref{eq:W_B}). This is achieved by invoking the following two procedures, which are also shared by conventional FFS:  (i) a \emph{basin simulator} routine  that propagates a long (MC or MD) trajectory from a configuration in $A$, and records all crossings of $\lambda_0$ (Algorithm~\ref{alg:basin-sim}), and (ii) an \emph{FFS iterator} that randomly chooses a configuration $x$ from a set of configurations in $\mathfrak{C}_k$, and propagates a trajectory until it crosses $\lambda_{k+1}$ or returns to $A$ (Algorithm~\ref{alg:ffs-it}).

We explain the operational similarities and differences between conventional FFS and jFFS by describing how a jFFS calculation is conducted. First, the basin simulator routine (Algorithm~\ref{alg:basin-sim}) takes as input a configuration $y_0\in A$, propagates a sufficiently long MD or MC trajectory $Y=(y_0,y_1,y_2,\cdots,y_m)$ from it, and identifies $y_j$'s at which $Y$ crosses $\lambda_0$ for the first time \emph{after} leaving $A$ at an earlier time. For each such $y_j$, the landing index, $s(y_j)$, is determined, which, for a  jumpy $\lambda(\cdot)$, can take any value between $0$ and $N$. This enables us to enumerate $s_0, s_1, \cdots, s_N$, the number of crossings of $\lambda_0$ resulting in configurations with landing indices $0,1,\cdots N$, respectively. Such $s_q$'s can then be used for estimating $\langle U_{-1,q}\rangle_{\mathcal{E}_A}$, which is related to a quantity that we call \emph{immediate flux} $\Psi_{A\rightarrow q}$:
	\begin{eqnarray}
	 \Psi_{A\rightarrow q} = \frac{\langle U_{-1,q}\rangle_{\mathcal{E}_A}}{\langle T_B\rangle_{\mathcal{E}_A}} &=& \frac{s_q}{m\delta{t}}
	\end{eqnarray}
	Here, $\delta{t}$ is the MD time step (or an equivalent MC sweep). For a smooth order parameter, all such crossings result in a configuration in $\mathfrak{C}_0$ and $s_1=s_2=\cdots=s_N=0$. Therefore all the immediate fluxes vanish except for $\Psi_{A\rightarrow 0}$, which is denoted by $\Phi_0$ in conventional FFS.
	
	In the next stage of jFFS, the basin simulator calls the FFS iterator routine (Algorithm~\ref{alg:ffs-it}) to pass along configurations corresponding to $\lambda_0$ crossings.  For each $0\le q<N$, all such $y_j$'s with landing index $q$ are passed along to an FFS iterator aimed at crossing  $\lambda_{q+1}$. Such an iterator  generates $N_t$ trial trajectories, and terminates them upon crossing $\lambda_{q+1}$ or returning to $A$. Unlike conventional FFS in which configurations at $\lambda_0$ are passed along to an iterator aimed at crossing $\lambda_1$, up to $N$ iterators can be called from within the basin simulator. For a jumpy order parameter, crossing $\lambda_{q+1}$ will result in $s_{q+1},s_{q+2},\cdots,s_N$ configurations with landing indices $q+1,q+2,\cdots,N$, unlike a smooth order parameter for which only $s_{q+1}\neq0$ and all resulting configurations fall at (or very close to) $\lambda_{q+1}$.

The $s_r~(r>q)$ configurations with a shared landing index $r$ are, in turn, sent to  an FFS iterator aimed at crossing $\lambda_{r+1}$. This recursive approach is necessary since, as outlined in Section~\ref{section:derive}, $\langle U_{i,j}\rangle_{\mathcal{E}_A}$ is history-dependent. Therefore, the configurations that share the same landing index-- or even the same $\lambda_q$, but arise from a different set of FFS iterations-- cannot be mixed and matched into a single FFS iterator. This implies that the FFS iterator routine of Algorithm~\ref{alg:ffs-it} can be called up to $f_N^{\text{jFFS}}=2^N-1$ times in jFFS, which is considerably larger than the $f_N^{\text{cFFS}}=N$ times that it is called in conventional FFS. The $2^N$th term in Eq.~(\ref{eq:W_B}) corresponds to a direct jump to $\mathfrak{C}_N=B$ upon crossing $\lambda_0$. $\langle U_{q,r}|\text{jump history}\rangle_{\mathcal{E}_A}$ can therefore be estimated from an FFS iterator aimed at crossing $\lambda_{q+1}$ as:
\begin{eqnarray}
\langle U_{q,r}|\text{jump history}\rangle_{\mathcal{E}_A} &=& \frac{s_r}{N_t}
\end{eqnarray}
and the overall rate is given by:
\begin{eqnarray}
\Phi_{A\rightarrow B}^{\text{jumpy}} &=& \Psi_{A\rightarrow N}+\sum_{q=0}^{N-1}\Psi_{A\rightarrow q}\sum_{k=1}^{N-q-1}\sum_{q<j_1<\cdots<j_k<N}\prod_{s=1}^{k+1} \langle U_{j_{s-1},j_s}|[-1,q,j_1,\cdots,j_{s-1}]\rangle_{\mathcal{E}_A}
\label{eq:rate-total-jffs}
\end{eqnarray}
One can similarly obtain $\Phi_{A\rightarrow i}$, the flux of trajectories that cross $\lambda_i$ after leaving $A$, and an associated cumulative transition probability from:
\begin{eqnarray}
\Phi_{A\rightarrow i}^{\text{jumpy}} &=& \sum_{j=i}^N\Psi_{A\rightarrow j}+\sum_{q=0}^{i-1}\Psi_{A\rightarrow q}\sum_{k=1}^{i-q-1}\sum_{q<j_1<\cdots<j_k<i\le j_{k+1}\le N}\prod_{s=1}^{k+1}\langle U_{j_{s-1},j_s}|[-1,q,j_1,\cdots,j_{s-1}]\rangle_{\mathcal{E}_A}\notag\\&&
\label{eq:psi-A-i}\\
P(\lambda_i|\lambda_0) &=& \frac{\Phi_{A\rightarrow i}^{\text{jumpy}}}{\Phi_{A\rightarrow 0}^{\text{jumpy}}}
\label{eq:jFFS-cumulative-prob}
\end{eqnarray}
Note the distinction between $\Phi_{A\rightarrow i}$ and $\Psi_{A\rightarrow i}$, as the former refers to total flux, while the latter corresponds to the flux of trajectories that immediately reach $\mathfrak{C}_i$ after crossing $\lambda_0$. Incidentally, $\Phi_{A\rightarrow 0}=\Psi_{A\rightarrow 0}$. 

The next question is to determine the statistical uncertainty in $\Phi_{A\rightarrow i}$.
For each non-vanishing pathway, the statistical uncertainty can be estimated using the approach described in Ref.~\cite{AllenFrenkel2006}. It is, however, necessary to emphasize that different jump pathways are not independent, and accounting for correlations between them is not straightforward. An upper bound can, however, be obtained for the error bar in $\langle W_B\rangle_A$ by adding up $\sigma_{j_0,j_1,\cdots,j_k,N}$'s, i.e.,~the uncertainties for individual non-vanishing jump pathways.

One further practical matter that makes the utilization of jFFS difficult is the potentially large number of FFS iterations. For easy tracking and tabulation of such iterations, we propose to map every iteration to an $N$-digit binary code, $B=b_1b_2\cdots b_N$, as follows. For an FFS iteration that starts at $\mathfrak{C}_p$ from configurations with the jump history $[-1,j_0,j_1,\cdots,j_k=p]$, $b_i$ will be given by $b_i=\delta_{i-1,j_1}+\delta_{i-1,j_2}+\cdots+\delta_{i-1,j_k}$. For instance, for an iteration starting at $\mathfrak{C}_5$  from configurations with jump history $[-1,1,2,4,5]$, $B=011011\underset{{N-6}}{\underbrace{0\cdots0}}$.

\subsubsection{Reduction of Iteration Count}
At first glance, it seems fairly complicated and computationally demanding to utilize jFFS, particularly due to exponential scaling of $f_N$ with $N$, the number of milestones. In practice, however, $\lambda^{\jmath}_m:=\langle|\lambda(x_{n+1})-\lambda(x_n)|\rangle_{n,x_0\in A}$ or the expected amount by which a jumpy order parameter can change per time step is not usually very large.  Therefore, $\lambda_i$'s can usually be chosen so that only jumps of one (or at most a few) milestones are possible in order to ensure that $f_N^{\text{jFFS}}$ scales linearly-- and not exponentially-- with $N$. In particular, if $\lambda_k$'s are chosen so that $\lambda_{k+1}-\lambda_k$ is always considerably larger than $\lambda^{\jmath}_m$, $f_N^{\text{jFFS}}\approx N$, and the only difference between conventional FFS and jFFS will be in using all configurations in $\mathfrak{C}_k$ sharing a common jump history-- and not only the ones at $\lambda_k$ irrespective of their jump history-- for estimating the probability of crossing $\lambda_{k+1}$. 

It might, however, still be the case that more than $N$ iterations might be needed if a calculation is conducted with a fixed pre-determined set of milestones. However, if $\lambda_i$'s are decided on-the-fly, i.e.,~if each $\lambda_{k+1}$ is decided after concluding the iteration aimed at crossing $\lambda_k$, the following  procedure can be used to ensure that exactly $N$ FFS iterations are conducted. For each iteration aimed at crossing $\lambda_k$, determine $\hat{\lambda}_k^{\max}$, or the largest value of the order parameter taken by a configuration obtained immediately after a first crossing of $\lambda_k$, and set $\lambda_{k+1}$ to be larger than $\hat{\lambda}_k^{\max}$. By doing this, it is assured that $s_{k+1}=s_{k+2}=\cdots=s_N=0$ within the routine aimed at crossing $\lambda_k$, and only one higher-order iteration will be called within each iteration. 

There are, however, situations at which the above-mentioned procedure is not practical, e.g.,~due to exceedingly small transition probabilities between $\lambda_k$ and $\hat{\lambda}_k^{\max}$. Even then, it can still be argued that in each iteration aimed at crossing $\lambda_k$, $s_k$ will almost always be considerably larger than $s_{k+1},s_{k+2},\cdots$. Therefore, the flux arising from the set of iterations  $0\rightarrow1\rightarrow\cdots\rightarrow N$ is  expected to be the largest contribution to the overall rate in Eq.~(\ref{eq:rate-total-jffs}). We have specifically formulated Algorithms~\ref{alg:basin-sim} and~\ref{alg:ffs-it} so that this most likely set of iterations, which we call the \emph{regular pathway} are called first. This would allow one to prematurely terminate a jump pathway if its partial cumulative flux is significantly smaller than the total flux of the regular pathway. Indeed, our numerical tests reveal that the cumulative fluxes of most jump pathways are several orders of magnitude smaller than that of the regular pathway, and the underestimation of rate in conventional FFS primarily arises from excluding the configurations that are not at (or very close to) $\lambda_k$ within an iteration aimed at crossing $\lambda_{k+1}$.

\begin{algorithm}
	\caption{\label{alg:basin-sim}Basin simulator}
   \begin{algorithmic}[1]
   	\State\textbf{Procedure} BasinSimulator
	\State For $y_0\in A$, generate a trajectory $Y\equiv(y_0,y_1,\cdots,y_m)$. \Comment{{\color{super-dark-red}Using MD or MC.}}
	\State $b:=1$. \Comment{{\color{super-dark-red}$y_0$ is in the $A$ basin.}}
	\For{$q=0,1,\cdots,N$} \Comment{{\color{super-dark-red}Loop over all milestones.}}
		\State$\mathcal{S}_q:=\{\}$. \Comment{{\color{super-dark-red}Empty config. list corresponding to landing in $\mathfrak{C}_q$ upon crossing $\lambda_0$.}}
		\State$s_q:=0$. \Comment{{\color{super-dark-red}Zero the counter corresponding to landing in $\mathfrak{C}_q$ upon crossing $\lambda_0$.}}
	\EndFor
	\For{$j=1,2,\cdots,m$} \Comment{{\color{super-dark-red} Analyze the trajectory $Y$ for crossing events.}
		\If {$y_j\in A$}}
			\State $b:=1$. \Comment{{\color{super-dark-red}Trajectory has returned to the $A$ basin.}}
		\EndIf
		\If {$\lambda(y_j)\ge\lambda_0$ and $b=1$}
			\State $b:=0$. \Comment{{\color{super-dark-red}Trajectory has not returned to $A$ after this crossing.}}
			\For{$l=0,1,\cdots,N$} \Comment{{\color{super-dark-red}Loop over milestones to determine the landing set.}}
				\If{$y_j\in\mathfrak{C}_l$} \Comment{{\color{super-dark-red}$\mathfrak{C}_l$ is the landing set.}}
					\State Add $y_j$ to $\mathcal{S}_l$.
					\State $s_l:=s_l+1$.
				\EndIf
			\EndFor
		\EndIf 
	\EndFor
	\For{$l=0,1,\cdots,N-1$}
		\If{$s_l>0$}
			\State\textbf{Call} FFSIterator($l,N_t,\mathcal{S}_l$). \Comment{{\color{super-dark-red}Send configurations landing in $\mathfrak{C}_l$ upon crossing $\lambda_0$ to FFSIterator.}}
		\EndIf
	\EndFor
	\State\Return $s_0,s_1,\cdots,s_N$.
   \end{algorithmic}
\end{algorithm}

\begin{algorithm}
	\caption{\label{alg:ffs-it}FFS iterator}
	\begin{algorithmic}[1]
		\State \textbf{Procedure} FFSIterator($k, N_t, \mathcal{C}=\{x_q\in\mathfrak{C}_k\})$.
		\Comment{{\color{super-dark-red}Conducts FFS iteration on configs in $\mathcal{C}$ all residing in $\mathfrak{C}_k$ by firing $N_t$ trial trajectories.}}
		\For{$l=k+1,\cdots,N$} \Comment{{\color{super-dark-red}Loop all milestones beyond $\lambda_k$.}}
			\State $\mathcal{S}_l:=\{\}$. \Comment{{\color{super-dark-red}Empty config. list corresponding to landing in $\mathfrak{C}_q$ upon crossing $\lambda_{k+1}$.}}
			\State $s_l:=0$. \Comment{{\color{super-dark-red}Zero the counter corresponding to landing in $\mathfrak{C}_q$ upon crossing $\lambda_{k+1}$.}}
		\EndFor
		\For{$i=1,2,\cdots,N_t$} \Comment{{\color{super-dark-red} Shoot $N_t$ trial trajectories.}}
			\State Randomly pick a configuration $y\in\mathcal{C}$.
			\State Launch a new trajectory $Y\equiv(y,y_1,y_2,\cdots)$ by randomizing momenta or choosing a new MC seed.
			\State Terminate $Y$ at $y_q$ if $\lambda(y_q)\ge\lambda_{k+1}$ or $y_q\in A$. \Comment{{\color{super-dark-red} $Y$ has crossed $\lambda_{k+1}$.}}
			\For{$l=k+1,k+2,\cdots,N$} \Comment{{\color{super-dark-red} Loop over remaining milestones to identify the landing set.}}
			\If {$y_q\in\mathfrak{C}_l$}  \Comment{{\color{super-dark-red} The crossing has resulted in a configuration in $\mathfrak{C}_k$.}}
				\State Add $y_q$ to $\mathcal{S}_l$. \Comment{{\color{super-dark-red} Add $y_q$ to the landing set.}}
				\State $s_l:=s_l+1$. \Comment{{\color{super-dark-red} Update the landing counter.}}
			\EndIf
			\EndFor
		\EndFor
		\For{$l=k+1,\cdots,N-1$}
			\If{$s_l>0$} \Comment{{\color{super-dark-red} Landing set $\mathcal{S}_l$ is not empty.}}
				\State \textbf{Call} FFSIterator ($l,N_{t,l},\mathcal{S}_l$). \Comment{{\color{super-dark-red} Conduct FFS iteration with config. set $\mathcal{S}_l$ aimed at crossing $\lambda_{l+1}$.}}
			\EndIf 
		\EndFor
		\State\Return{$s_{k+1},s_{k+2},\cdots,s_N$.}
	\end{algorithmic}
\end{algorithm}

\begin{figure}
\centering
\includegraphics[width=.43\textwidth]{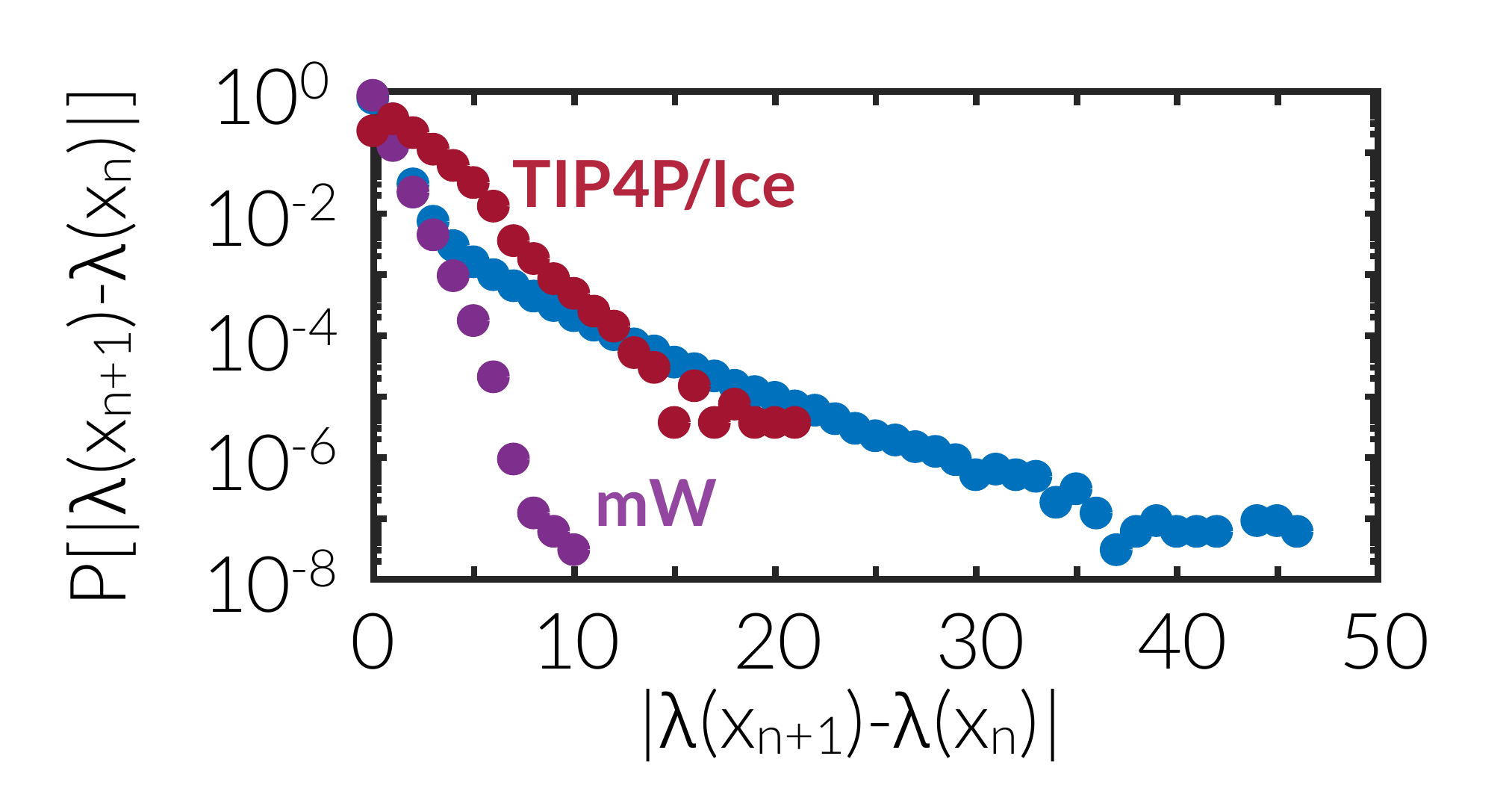}
\caption{\label{fig:op-jump}
$P(|\lambda(x_{n+1})-\lambda(x_n)|)$ vs.~$|\lambda(x_{n+1})-\lambda(x_n)|$ for the Lennard-Jones, mW and TIP4P/Ice systems. For the Lennard-Jones system, this distribution is computed with a sampling time of $t_s^*=0.0025$ at $T^*=0.47$ and $p^*=0$. For mW and TIP4P/Ice, the distributions are computed at $T=230$~K and $p=1$~bar, with  sampling times of 2~fs, and 1~ps, respectively.
}
\end{figure}

%\begin{figure}
%\centering
%\includegraphics[width=.4\textwidth]{FFS-LJ.pdf}
%\caption{\label{fig:ffs-lj} Deviations of the rates computed from conventional FFS and MFPT from the rate computed from jFFS in homogeneous crystal nucleation in the Lennard-Jones liquid at $p^*=0$.  }
%\end{figure}
%

\section{Simulation Methodology}
\label{section:methods}

In order to assess the extent by which the rate of a rare event described by a jumpy order parameter is underestimated upon using conventional FFS, we compute homogeneous crystal nucleation rates using both conventional and jumpy FFS in the following systems: (i) the Lennard-Jones~\cite{LJProcRSoc1924} system at zero pressure, (ii) the monoatomic water (mW)~\cite{MolineroJPCB2009} system at 1~bar, (iii) the TIP4P/Ice~\cite{VegaTIP4PiceJCP2005} system (a molecular model of water) at 230~K and 1~bar.
The order parameter utilized in each system will be described in Section~\ref{section:order-parameter}.
 As can be seen in Fig.~\ref{fig:op-jump}, however, the utilized order parameters are all jumpy for these three systems, since jumps of larger than $\pm1$ are very likely in all systems. For the first two systems, we also compute the rates using the mean free passage time (MFPT) method~\cite{WedekindJChemPhys2007} whenever possible. This latter method is based on analyzing unbiased crystallizing MD trajectories, and therefore enables us to compare the rates computed from FFS and jFFS with the actual rates extracted from unbiased MD simulations.  For each state point, our MFPT analysis is based on a minimum of 75 independent unbiased MD trajectories. The calculation in (iii) is a partial repeat of our earlier calculation reported in Ref.~\cite{HajiAkbariPNAS2015} using jFFS, as it is not practical to conduct the  calculation at its entirety  due to its prohibitively large computational cost.

\subsection{Molecular Dynamics Simulations}
All molecular dynamics simulations are conducted using LAMMPS~\cite{PimptonLAMMPS1995}. Newton's equations of motion are integrated using velocity-Verlet algorithm~\cite{SwopeJCP1982} with  time steps of $\Delta t^*=0.0025$ for the LJ system and $\Delta t=2~$fs for the mW and TIP4P/Ice systems, respectively. All simulations are conducted in the $NpT$ ensemble, with temperature and pressure controlled using the Nos\'{e}-Hoover thermostat~\cite{NoseMolPhys1984, HooverPhysRevA1985} and Parrinelo-Rahman barostat~\cite{ParrinelloJAppPhys1981}. For each simulation, the time constants for the thermostat and the barostat are chosen as $\tau_{\text{thermostat}}=10^2\Delta t$ and $\tau_{\text{barostat}}=10^3\Delta t$, respectively. In the TIP4P/Ice system, long-range electrostatic interactions are treated using the particle-particle particle-mesh (PPPM) method~\cite{Hockney1989}, with a cutoff of 0.85~nm for the short-range part. We also apply the SHAKE algorithm to enforce the rigidity of water molecules~\cite{RyckaertJCompPhys1977}. 

\noindent
\subsection{System Preparation}
For the LJ and mW systems, initial configurations are obtained from melting each system's respective crystal (FCC for LJ and cubic ice for mW) at a sufficiently high temperature, and gradually quenching the arising configurations to the target temperature and pressure.
The initial crystalline configurations are comprised of 6192 atoms in the case of LJ and 4096 atoms in the case of mW and are melted at $T^*=1$ and $T=350$~K, respectively. For the TIP4P/Ice system, no new configurations are generated. Instead, we utilize the basin configurations obtained in our earlier calculation presented in Ref.~\cite{HajiAkbariPNAS2015}, with the preparation process thoroughly explained therein. 

\subsection{Order Parameter}\label{section:order-parameter}
In studies of crystal nucleation, the order parameter is typically chosen as the number of atoms and/or molecules in the largest crystalline nucleus in the system. First, each atom or molecule is classified as solid-like or liquid-like based on its local environment. The neighboring solid-like atoms (molecules) are then clustered together to form crystalline nuclei of different sizes. In all the systems considered in this work, the identity of each atom or molecule is determined using Steinhardt bond-order parameters~\cite{SteinhardtPRB1983} with details presented in our earlier publications~\cite{HajiAkbariFilmMolinero2014, HajiAkbariPNAS2015}. In summary, for  atom (or molecule) $r$ in the mW and TIP4P/Ice system, $q_l(r)$ is computed as:
\begin{eqnarray}
q_l(r) &=& \frac{1}{N_b(r)}\sum_{s=1}^{N_b(r)}\frac{\textbf{q}_l(r)\cdot\textbf{q}_l^*(s)}{|\textbf{q}_l(r)||\textbf{q}_l(s)|}
\end{eqnarray}
with $\textbf{q}_l\equiv(q_{l,-l},q_{l,-l+1},\cdots,q_{l,l})$  a vector in $\mathbb{C}^{2l+1}$, and $N_b(r)$, the number of atoms (molecules) that are in the first nearest neighbor shell of molecule $r$, i.e.,~are within a distance of $r_c=0.32$~nm from $r$. Here, $\textbf{q}_l(r)\cdot\textbf{q}_l^*(s) = \sum_{m=-l}^{l}q_{lm}(r)q_{lm}^*(s)$ is the inner product between $\textbf{q}_l(r)$ and that of its $s$th neighbor. The components of $\textbf{q}_l(r)$ are given by:
\begin{eqnarray}
q_{lm}(r) &=& \frac{1}{N_b(r)} \sum_{s=1}^{N_b(r)} Y_{lm}(\theta_{rs},\phi_{rs}),~~-l\le m\le l
\end{eqnarray}
Here, $\theta_{rs}$ and $\phi_{rs}$ are spherical angles associated with the displacement vector, $\textbf{r}_{ij}=\textbf{r}_j-\textbf{r}_i$, and $Y_{lm}$'s are spherical harmonics functions. Any molecule that has a $q_6(r)\ge0.5$ is classified as solid-like. The neighboring solid-like molecules are then clustered, with the arising clusters further refined using the chain exclusion algorithm of Ref.~\cite{VegaJCP2012}. In the TIP4P/Ice system, all calculations are conducted based on the positions of oxygen atoms.  In the LJ system, a cutoff  of $r_c=1.41\sigma$ is utilized~\cite{DelagoJCP2008}, and for each atom, a neighbor-averaged $\overline{\textbf{q}}_6$ is computed as:
\begin{eqnarray}
\overline{\textbf{q}}_6(r) &=& \frac{1}{N_b(r)}\sum_{s=0}^{N_b(r)} \textbf{q}_6(s)\\
\overline{q}_6(r) &=& \sqrt{
\frac{4\pi}{13}\sum_{m=-6}^6|\overline{q}_{6m}(r)|^2
}
\end{eqnarray}
If $\bar{q}_{6}(r)\ge0.3$, atom $r$ is classified as solid-like. No chain exclusion algorithm is applied to the clusters obtained in the LJ system.

\subsection{FFS Iterations}
We conduct conventional FFS, jFFS and MFPT using \texttt{AdvSamp}, our in-house C++ trajectory manager program that links against LAMMPS, a package also used and discussed in further details in our earlier publications~\cite{GianettiPCCP2016, HajiAkbariPNAS2017}. As mentioned in Section~\ref{section:mathematical}, FFS iterations conducted in jFFS are classified into  two categories, based on the jump history of starting configurations. An FFS iteration starting from $\mathcal{C}$, a set of configuration in $\mathfrak{C}_k$, is called \emph{regular} if every $x\in\mathcal{C}$ has the jump history, $[-1,0,1,\cdots,k]$, and is called \emph{non-regular} otherwise. For the LJ and mW systems, regular iterations are terminated after a minimum of 1,000 crossings, with more crossings (2,000-3,000) required for the first few milestones. For the TIP4P/Ice system, however, regular iterations are terminated after a minimum of 500 crossings, with 2,000 crossing required for the first four milestones. As mentioned earlier, multi-milestone jumps, i.e.,~situations in which $\langle U_{i,j}\rangle\neq0$ for $j>i+1$, are far less common than regular crossings (from $\mathfrak{C}_i\rightarrow\mathfrak{C}_{i+1}$). This means that usually, fewer starting configurations are available for a non-regular iteration. Let the number of starting configurations and trial trajectories of the regular iteration starting from $\mathfrak{C}_k$ be $N_c$ and $N_t$, respectively. For a non-regular iteration starting from $N_c'$ configurations in $\mathfrak{C}_k$, we use a minimum of $N_t'=N_c'N_t/N_c$ trial trajectories. In other words, we make sure that the number of trial trajectories initiated per configuration in a non-regular iteration is as large as that of the regular iterations starting from the same milestone. This choice is made to assure that the computational cost of a jFFS calculation is kept reasonable, by avoiding unnecessary integration of a large number of trajectories from a handful of configurations usually available for a non-regular iteration.

\section{Results and Discussions}
\label{section:results}
\begin{figure*}
	\centering
	\includegraphics[width=.82\textwidth]{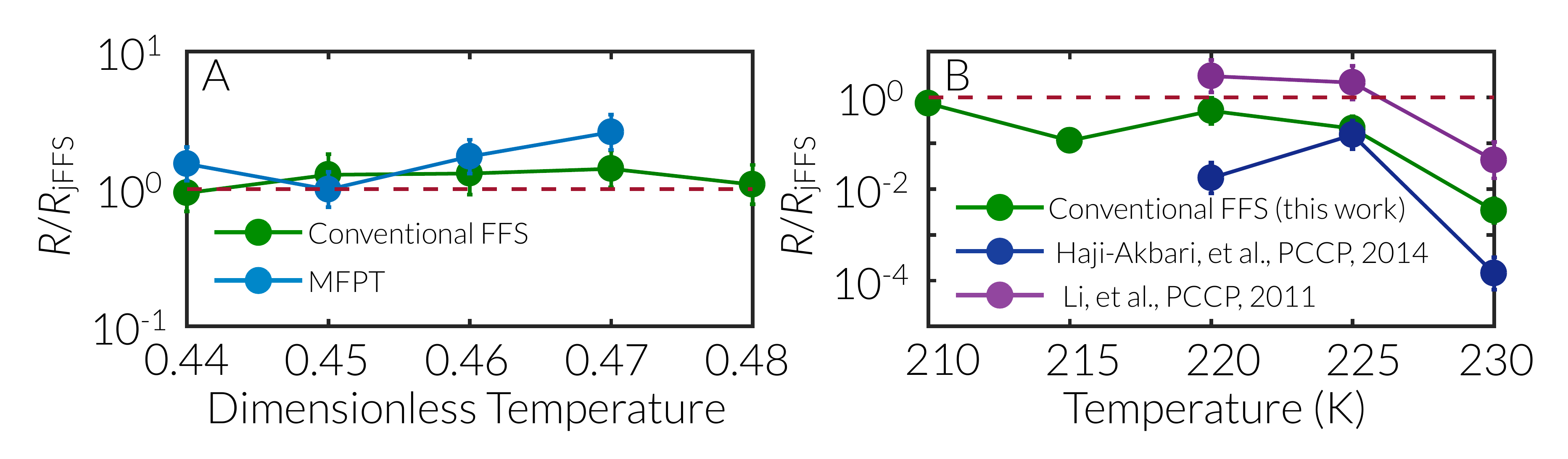}
	\caption{\label{fig:ffs-lj-mw} Deviations of the homogeneous crystal nucleation rates computed from conventional FFS and MFPT from those estimated through jFFS in (A) the LJ system at $p^*=0$, and (B) the mW system at $p=1$~bar. In the mW system, MFPT calculations are conducted at one temperature only. The literature data in (B) are from Refs.~\cite{GalliPCCP2011, HajiAkbariFilmMolinero2014}.
	}
\end{figure*}

Figs.~\ref{fig:ffs-lj-mw}A and~\ref{fig:ffs-lj-mw}B show $R/R_{\text{jFFS}}$ vs.~temperature in the LJ and mW systems, respectively. Here, $R$ is the volumetric nucleation rate computed using MFPT and conventional FFS. All absolute rates are given in Tables~\ref{table:rates-lj} and \ref{table:rates-mw}. For the mW system, we also include the rates reported in Refs.~\cite{GalliPCCP2011, HajiAkbariFilmMolinero2014}. The observed discrepancy between conventional FFS and jFFS is statistically insignificant in the LJ system. In the mW system, however, conventional FFS underestimates rates by up to four orders of magnitude. We suspect that this qualitative difference is due to larger cumulative probabilities-- as defined by (\ref{eq:jFFS-cumulative-prob})-- in the LJ system  ($\approx10^{-8}$ for the LJ system at $T^*=0.48$ vs.~$\approx10^{-20}$ for mW system at T=230~K) even though we cannot rule out the possibility that the difference could arise from a deeper physical difference between the two systems. A more thorough investigation of this issue can be the topic of future studies. 

In addition to discrepancies between the cFFS and jFFS rate estimates in the mW system, we observe a modest-- but statistically significant-- discrepancy between cFFS rates computed here, with the rates reported in Refs.~\cite{GalliPCCP2011, HajiAkbariFilmMolinero2014}. These discrepancies seem odd considering that conventional FFS is expected to be not very sensitive to technical details such as the order parameter, and positioning of milestones. But one should note that this lack of sensitivity is predicated upon the proximity of the utilized order parameter to the true reaction coordinate, as well as proper sampling of the reactive trajectory ensemble. Even though there is good evidence that the order parameters utilized here and in Refs.~\cite{GalliPCCP2011, HajiAkbariFilmMolinero2014} are reasonably close to the reaction coordinate for homogeneous nucleation~\cite{LupiNature2017}, undersampling of the reactive trajectory ensemble by cFFS is potentially sufficient to make it susceptible to details such as the order parameter, and the positioning of milestones. 

%As for the mW system, the discrepancy between the nucleation rates reported in Refs.~\cite{GalliPCCP2011, HajiAkbariFilmMolinero2014} can be attributed to the possible sensitivity of conventional FFS to details of the order parameter and positioning of milestones, as Refs.~\cite{GalliPCCP2011, HajiAkbariFilmMolinero2014} employ slightly different order parameters. In other words, even though conventional FFS should, in principle, not be very sensitive to the particular choice of an order parameter and positioning of milestones, its inability to accurately predict the rate for a jumpy order parameter can make it potentially susceptible to such technical details. 

\begin{table*}
	\centering
	\caption{\label{table:rates-lj}Nucleation rates computed in the Lennard-Jones system at $p^*=0$}
	\begin{tabular}{ccccccc}
	\hline\hline
	$T$ && $\log_{10}R_{\text{MFPT}}$ && $\log_{10}R_{\text{cFFS}}$ && $\log_{10}R_{\text{jFFS}}$ \\
	\hline
	0.44 && $-6.8103\pm0.0919$ && $-7.0218\pm0.1084$ && $-6.9931\pm0.0810$ \\
	0.45 && $-7.8231\pm0.0883$ && $-7.7156\pm0.1189$ && $-7.8193\pm0.0930$  \\
	0.46 && $-9.1557\pm0.0734$&& $-9.2788\pm0.1178$ && $-9.3918\pm0.0986$ \\
	0.47 && $-10.7237\pm0.0883$ && $-10.9905\pm0.0982$ && $-11.1376\pm0.0927$\\
	0.48 &&- && $-13.3270\pm0.1023$ && $-13.3593\pm0.0998$ \\
	\hline
	\end{tabular}
\end{table*}

\begin{table*}
	\centering
	\caption{\label{table:rates-mw}Nucleation rates computed in the mW system at $p=1~$atm. Nucleation rates are in $\text{m}^{-3}\cdot\text{s}^{-1}$.}
	\begin{tabular}{ccccccc}
	\hline\hline
	$T$~(K) && $\log_{10}R_{\text{MFPT}}$ && $\log_{10}R_{\text{cFFS}}$ && $\log_{10}R_{\text{jFFS}}$ \\
	\hline
	210 & & $+31.3738\pm0.0992$ & & $+31.3849\pm0.1023$ & &  $+31.5171\pm0.0807$\\
	215 & & - & & $+28.5597\pm0.1428$ & & $+29.5037\pm0.1132$ \\
	220 & & - & & $+24.5743\pm0.1328$ & & $+24.8672\pm0.2454$ \\
	225 & & - & & $+19.0375\pm0.1763$ & & $+19.7176\pm0.1978$ \\
	230 & & - & & $+12.5118\pm0.2099$ & & $+14.9825\pm0.1667$\\
	\hline
	\end{tabular}
\end{table*}

%\begin{figure}
%	\centering
%	\includegraphics[width=.5\textwidth]{Rates-mW.pdf}
%	\vspace{-20pt}
%	\caption{\label{fig:rates-mw}
%	Deviations of conventional FFS rate calculations from the rate obtained from jFFS in the mW system. The literature data are from Refs.~\cite{GalliPCCP2011, HajiAkbariFilmMolinero2014}. Only on calculation was conducted using MFPT at 210~K and 1~bar, with $\log_{10}[R_{\text{MFPT}}/R_{\text{jFFS}}]=-0.1433\pm0.1279$.
%	}
%\end{figure}

\begin{figure}
\centering
\includegraphics[width=.4\textwidth]{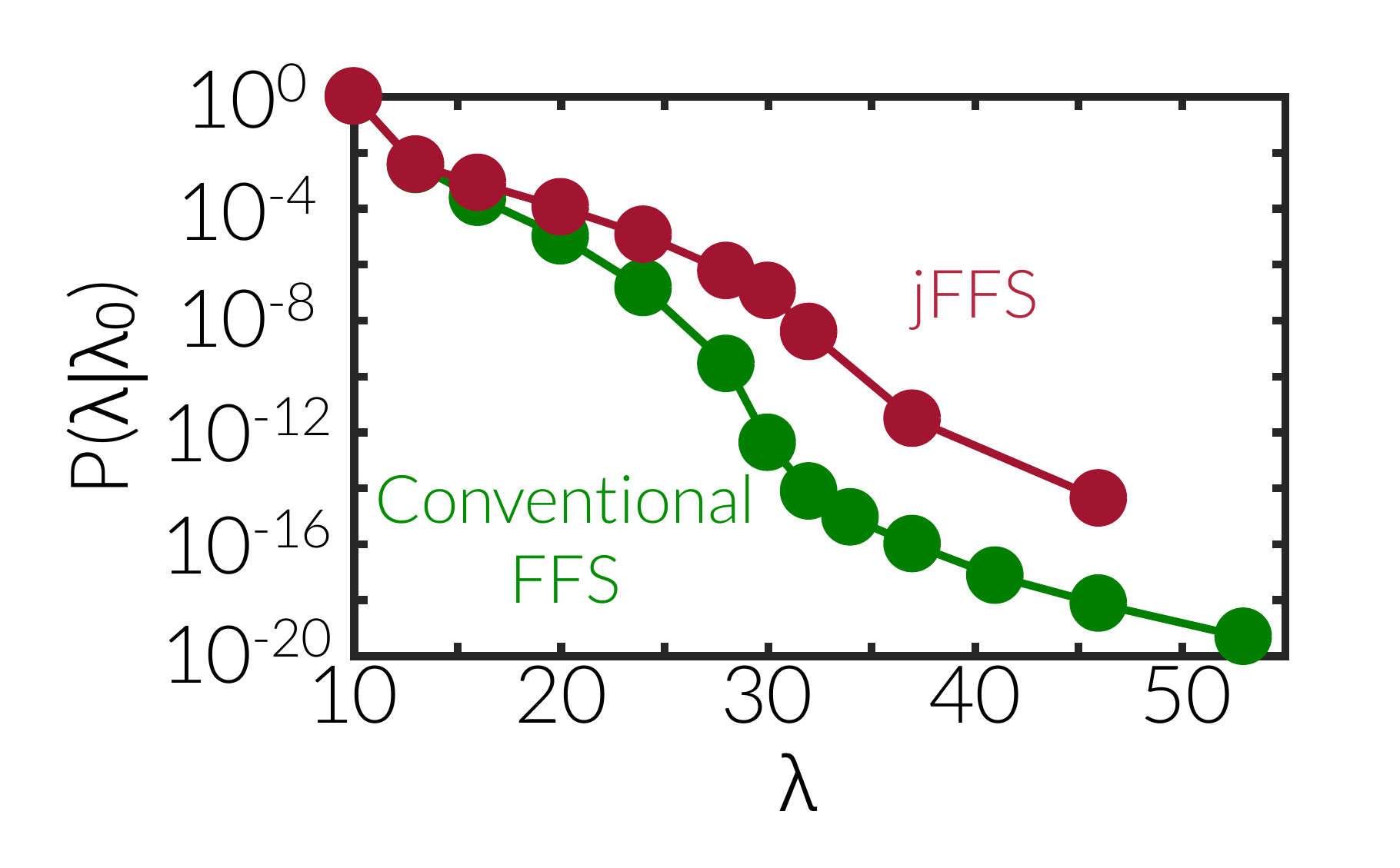}
\caption{\label{fig:partial-repeat}
Partial repeat of the rate calculation initially reported in Ref.~\cite{HajiAkbariPNAS2015}. For the old calculation, $P(\lambda|\lambda_0)$ is defined as described in Ref.~\cite{HajiAkbariPNAS2015}, while in the new calculation, $P(\lambda|\lambda_0)$ is obtained from Eq.~(\ref{eq:jFFS-cumulative-prob}). 
jFFS and conventional FFS yield values of $\log_{10}P_{\text{jFFS}}(46|\lambda_0=10)=-14.3759\pm0.3835$ and $\log_{10}P_{\text{FFS}}(46|\lambda_0=10)=-18.1365\pm0.1677$, respectively. Error bars are smaller in size than utilized symbols.
}
\end{figure}

\begin{figure}
	\includegraphics[width=.41\textwidth]{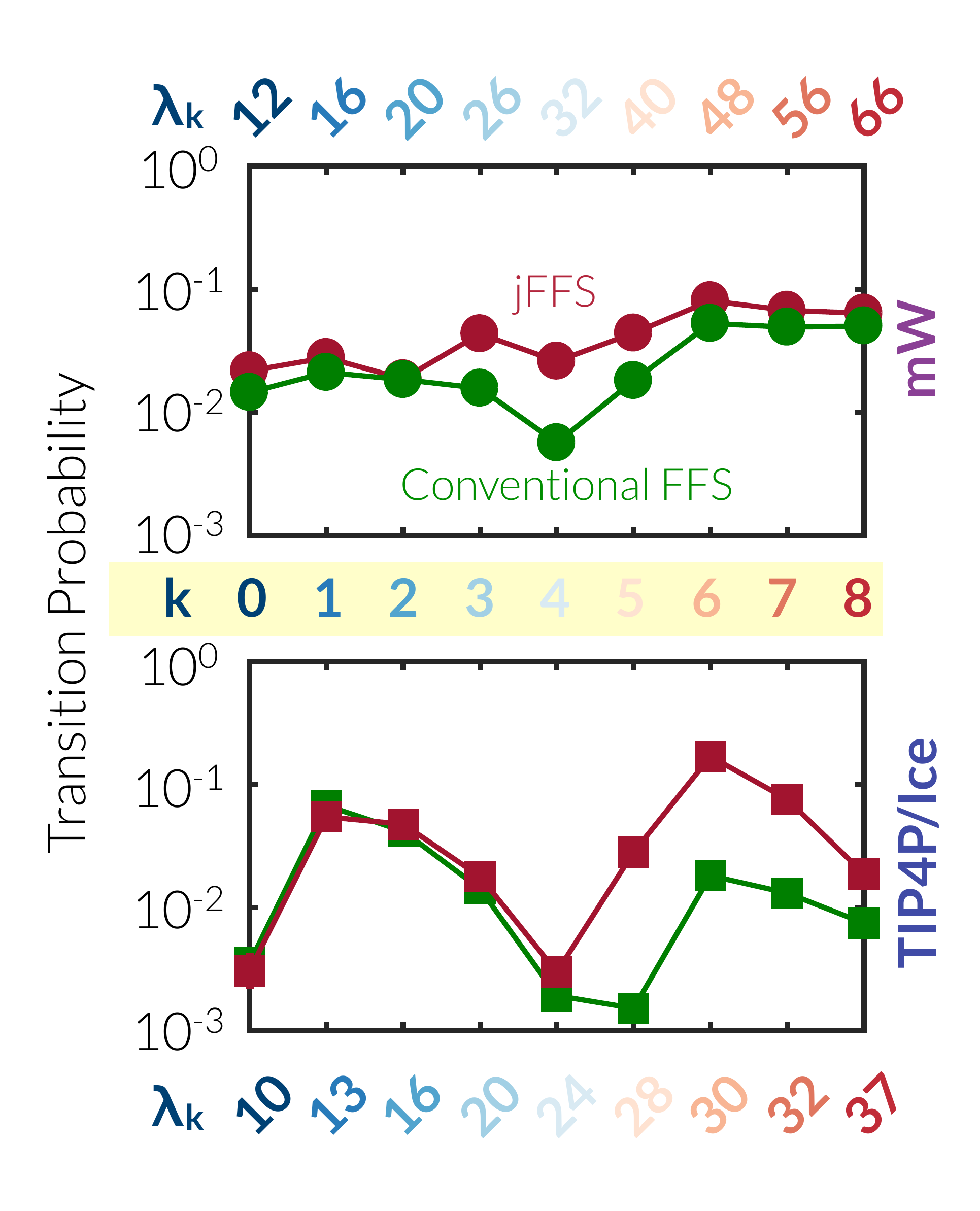}
	\caption{\label{fig:prob-compare}Early-milestone transition probabilities for the conventional FFS and jFFS calculations of homogeneous ice nucleation rate in the mW and TIP4P/Ice systems. Like Ref.~\cite{HajiAkbariPNAS2015}, a sampling time of 1~ps is used in the TIP4P/Ice calculation. For the jFFS calculation, transition probability corresponds to $\langle U_{k,k+1}\rangle_{[-1,0,\cdots,k-1,k]}$. Error bars are smaller in size than utilized symbols.}
\end{figure}

The largest difference between conventional FFS and jFFS is observed in the partial  repeat of our earlier calculation of homogeneous ice nucleation rate in the TIP4P/Ice system at 230~K and 1~bar~\cite{HajiAkbariPNAS2015}. Due to large computational costs, we  repeat this calculation partially, only through the inflection region reported in Ref.~\cite{HajiAkbariPNAS2015} and depicted in Fig.~\ref{fig:partial-repeat}, i.e.,~up to $\lambda_9=46$. 
%We observe that the pathways starting wth $-1\rightarrow1$ and $-1\rightarrow2$ do not reach $\lambda_9=46$, which enables us to compute an equivalent $P(\lambda|\lambda_0)$ by dividing $\Phi_{A\rightarrow\lambda}$ in Eq.~(\ref{eq:psi-A-i}) by $\Phi_{A\rightarrow0}$. 
As can be seen in Fig.~\ref{fig:partial-repeat}, applying jFFS results in  weakening of the inflection, and an increase in $P(\lambda_9=46|\lambda_0=10)$ by $+3.7606\pm0.4186$ orders of magnitude.  Apart from the inflection region where milestones are very close, $\langle U_{i,i+1}\rangle$'s do not differ significantly from the $P(\lambda_{i+1}|\lambda_i)$'s reported in the old calculation (Fig.~\ref{fig:prob-compare}). We therefore expect the total rate not to exceed the rate reported in Ref.~\cite{HajiAkbariPNAS2015} by more than four orders of magnitude, especially since the remaining milestones are too distant from one another for the jumpiness of $\lambda$ to be important. Similarly, deviations from jFFS in the mW system are also confined to small $\lambda$'s where milestones are relatively close (Fig.~\ref{fig:prob-compare}). 

It is necessary to emphasize that this calculation is conducted using coarse-grained FFS in which $\lambda$ is evaluated every 1~ps, and not at every MD step. As can be seen in Fig.~\ref{fig:op-jump}, this coarse-graining results in larger jumps in $\lambda(t)$-- in comparison to the mW system-- and leads to many multi-milestone jumps. For the 9-milestone calculation of Fig.~\ref{fig:partial-repeat} therefore, we conduct 63 FFS iterations to accurately determine history-dependent transition probabilities. Interestingly, four distinct jump pathways result in configurations in $\mathfrak{C}_9$, corresponding to cumulative log probabilities of  $-14.5451\pm0.2330, -14.9487\pm0.7113, -15.7220\pm0.3534$ and $-16.3678\pm2.5863$, respectively. Note that the second most likely pathway has a partial cumulative probability that is $39\%$ of that of the regular pathway. This is unlike the LJ and mW systems in which the contribution of such non-regular pathways to the overall rate never exceeds 2\% of the overall rate. It is necessary to emphasize that this calculation is a partial repeat of the full rate calculation reported in Ref.~\cite{HajiAkbariPNAS2015}, and we do not really know whether these non-regular pathways will survive-- let alone contribute significantly to the overall rate-- if further FFS iterations are conducted. However, the fact that multiple jump pathways can have comparable contributions to the partial flux is remarkable and demonstrates the potential peril in neglecting multi-milestone jumps in FFS.

\begin{figure}
	\includegraphics[width=.45\textwidth]{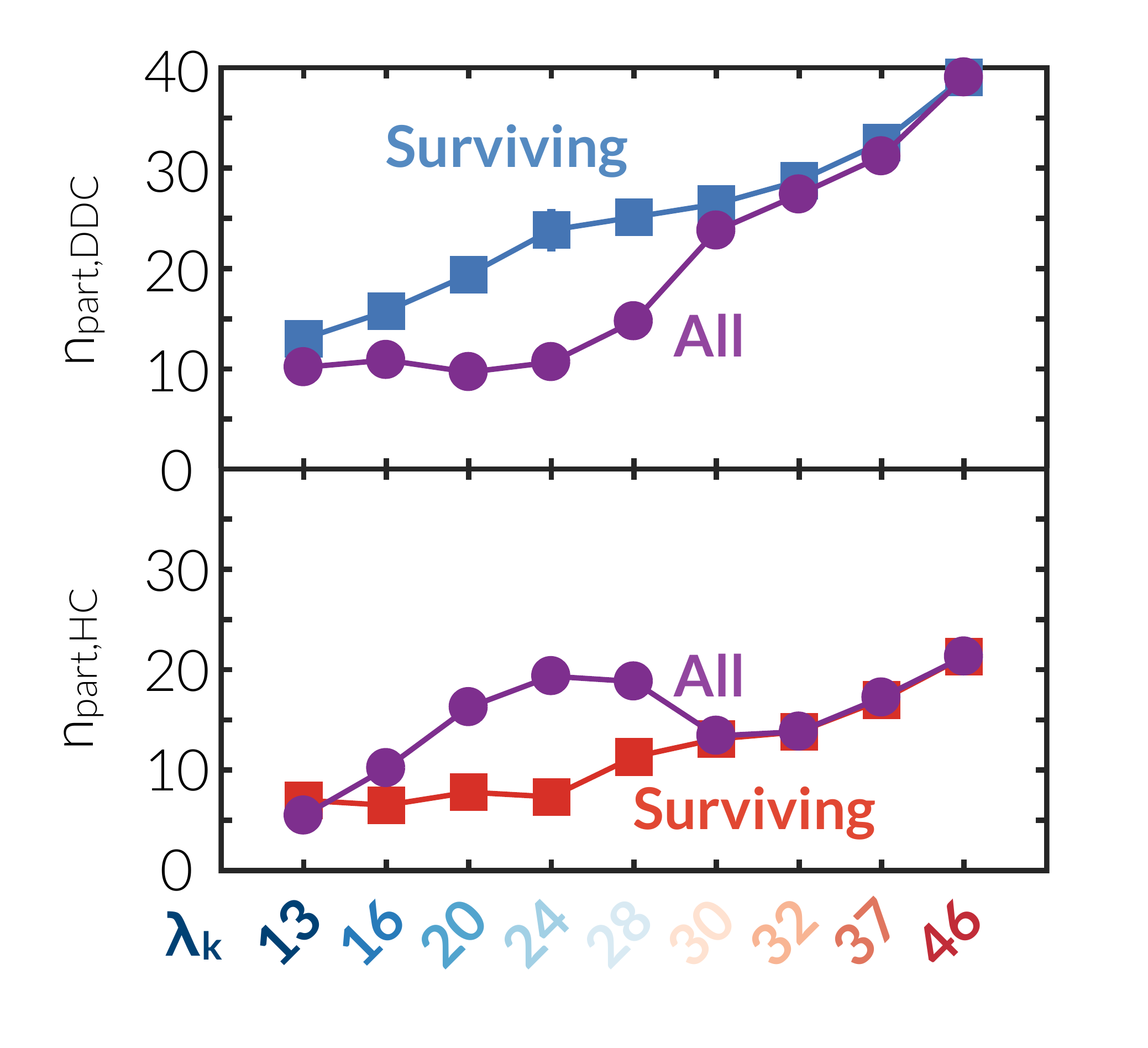}
	\caption{\label{fig:cages-ffs} Cage participation of the oxygen atoms that are part of the largest solid-like cluster. For each $\lambda_k$, all configurations in $\mathfrak{C}_k$ are considered, with the surviving configurations having progeny at $\mathfrak{C}_9=\{x\in\mathscr{Q}: \lambda(x)\ge46\}$. Note the dramatic difference between the cage participation of surviving configurations, which are significantly more cubic than all the configurations collected at each $\mathfrak{C}_k,k<9$. Error bars are smaller in size than utilized symbols.}
\end{figure}

In addition to predicting the nucleation rate, FFS can provide important mechanistic information about the underlying rare event. For instance, in Ref.~\cite{HajiAkbariPNAS2015}, we conducted a careful analysis of cage statistics to conclude that the inflection in $\lambda\approx30$ is due to competition between double-diamond cages (DDCs) and hexagonal cages (HCs), which are the topological building blocks of cubic and hexagonal ice, respectively. Since cFFS and jFFS predict widely different rates, it is important to determine whether they also reveal qualitatively different nucleation mechanisms.  We thus repeat the cage analysis conducted in Ref.~\cite{HajiAkbariPNAS2015}. In accordance with the pedigree analysis approach introduced in Ref.~\cite{HajiAkbariPNAS2015}, we trace back the ancestry of all configurations in $\mathfrak{C}_9$, to identify configurations at earlier $\mathfrak{C}_k$'s that have some offspring at $\mathfrak{C}_9$.  We then compute the average number of water molecules that are simultaneously part of the largest solid-like cluster, and a DCC or HC. Unlike the rate calculation, we do not distinguish between the configurations that are in the same $\mathfrak{C}_k$, but have different jump histories. 
As can be seen in Fig.~\ref{fig:cages-ffs}, the surviving configurations are significantly more cubic than an average configuration gathered at $\mathfrak{C}_k$. This is consistent with the mechanism proposed in Ref.~\cite{HajiAkbariPNAS2015}, and shows that using conventional FFS only results in an underestimation of  rate, while the observed nucleation mechanism remains qualitatively unchanged. 
We do not know whether this finding is general. However, it can be argued that using conventional FFS with a jumpy order parameter simply pins the reactive pseudo-trajectories to pass through artificially chosen milestones. Even though this will most likely result in quantitative discrepancies e.g.,~in cage participation in the TIP4P/Ice system, it is not expected to yield mechanisms that are qualitatively distinguishable from the actual mechanism. In other words, if the utilized order parameter is sufficiently close to the underlying reaction coordinate, such pinning will be akin to taking pictures of an event at artificially chosen positions, instead of all the way through the process. This will, however, be only true if the contribution of the regular pathway to the overall rate is significantly larger than non-regular pathways. For processes that are strongly affected by jumps in the order parameter, e.g.,~coalescence of precritical clusters, this might no longer be the case, and the mechanism inferred from conventional FFS might be markedly different from the actual mechanism.

\section{Conclusions}
\label{section:summary}
In summary, we develop a modified version of FFS for which the smoothness of the utilized order parameter and sequential crossing of FFS milestones is no longer necessary. We conduct numerical tests and use conventional FFS and jFFS to calculate the rate of homogeneous crystal nucleation in several systems. We conclude that using conventional FFS can result in considerable underestimation of the nucleation rate.  Considering the increased popularity of FFS in studying  rare event phenomena, this generalized method can be of broad interest to the computational statistical mechanics community. Furthermore, the proposed approach can guide future efforts in developing and generalizing milestone-based path sampling techniques, such as transition interface sampling~\cite{BolhuisJCompPhys2005}. Further studies are, however, necessary to develop jumpy extensions of other variants of FFS, such as the branched growth and the "Rosenbluth-like`` method~\cite{FrenkelFFS_JCP2006}, as well as the formalism to be utilized in extracting free energy surfaces from forward~\cite{ThaparJCP2015}, and forward and reverse jFFS calculations~\cite{ValerianiJChemPhsy2007}.

\begin{acknowledgments}
A.H.-A. gratefully acknowledge the support of the National Science Foundation CAREER Award (Grant No. CBET-1751971). A.H.-A. acknowledges useful discussions with P. G. Debenedetti and S. Sarupria. These calculations were performed on the Yale Center for Research Computing. This work used the Extreme Science and Engineering Discovery Environment (XSEDE), which is supported by National Science Foundation grant number ACI-1548562.
\end{acknowledgments}

\nocite{TownsCompSciEng2014}

\appendix

\section{\label{sec:Uij_derive}History Dependence of Transition Probabilities}

\noindent The probability density function for a discrete-time trajectory, $X\equiv(x_0,x_1,x_2,\cdots)$ is given by:
\begin{eqnarray}
\mathscr{P}_0(X) &=& C_0\rho_0(x_0)\prod_{q=0}^{\infty}\pi(x_{q+1}|x_q)
\end{eqnarray}
Here, $C_0$ is the normalization constant, and $\rho_0(\cdot)$ is the equilibrium distribution of configurations in $\mathscr{Q}$. Now, define $\xi_i^n(x_n)$ as:
\begin{eqnarray}
\xi_i^1(x_1) &:=& \int dx_0\rho_0(x_0)\theta_A(x_0)\pi(x_1|x_0)\notag\\
\xi_i^n(x_n) &:=& \int\Bigg\{\left[\prod_{q=0}^{n-1}dx_q\pi(x_{q+1}|x_q)\right]\rho_0(x_0)\sum_{a=1}^{n-1}\left[\prod_{q=0}^{a}\theta_A(x_q)\prod_{q=a+1}^{n-1}\phi_i(x_q)\right]\Bigg\}
\label{eq:xi_in}
\end{eqnarray}
Note that $\xi_i^n(x)$ is the probability that $X_n\equiv(x_0,x_1,\cdots,x_{n-1},x)$, a partial trajectory starting in $A$ and ending in $x$: (i) never returns to $A$ after possibly leaving it at some $x_k, (k<n)$, (ii) never crosses $\lambda_i$ before reaching $x$.  It is easy to note that  $\langle U_{i,j}\rangle$ can be expressed as:
\begin{eqnarray}
\langle U_{i,j}\rangle &=& \sum_{1\le a<b}\int dx_a^b\left[\prod_{q=a}^{b-1}\pi(x_{q+1}|x_q)\right]\xi_i^{a}(x_a)\theta_i(x_a)\theta_j(x_b)\notag\\&&\times\prod_{q=a+1}^{b-1}\phi_{i+1}(x_q)= \int dx\omega_i(x)\theta_j(x)
\end{eqnarray}
Here, $dx_a^b=\prod_{q=a}^bdx_q$ and $\omega_i(x)$ is given by:
\begin{eqnarray}
\omega_i(x) &:=& \sum_{b=2}^{\infty}\sum_{a=1}^{b-1}\int dx_a^{b-1}\left[\prod_{q=a}^{b-2}\pi(x_{q+1}|x_q)\right]\xi_i^a(x_a)\theta_i(x_a)\pi(x|x_{b-1})\prod_{q=a+1}^{b-1}\phi_{i+1}(x_q)\label{eq:omega_i}
\end{eqnarray}
$\omega_i(x)$ is the probability that a trajectory starting in $A$ and ending in $x$ crosses $\lambda_i$ into $\mathfrak{C}_{i}$ at some point in between, and never crosses $\lambda_{i+1}$ or returns to $A$ prior to reaching $x$. Note that $\omega_i(x)$ satisfies the following recursion:
\begin{eqnarray}
\omega_i(x) &=& \int dx_1\pi(x|x_1)\left\{\theta_i(x_1)\xi_i(x_1) + \phi_{i+1}(x_1)\omega_i(x_1)\right\}\notag\\&&
\end{eqnarray}
with $\xi_i(x)$ given by:
\begin{eqnarray}
\xi_i(x) &:=& \sum_{a=1}^{\infty}\xi_i^a(x)
\end{eqnarray}
Similarly, $\langle U_{i,j}U_{j,k}\rangle$ can be expressed as:
\begin{eqnarray}
\langle U_{i,j}U_{j,k}\rangle &=& \sum_{a<b<c} \int\left[\prod_{r=a}^cdx_r\right]\Bigg\{\xi_i^a(x_a)\theta_i(x_a)\notag\\
&&\times\left[\prod_{r=a+1}^{b-1}\pi(x_r|x_{r-1})\phi_{i+1}(x_r)\right]\theta_j(x_b)\pi(x_b|x_{b-1})\notag\\
&&\times\left[\prod_{r=b+1}^{c-1}\pi(x_r|x_{r-1})\phi_{j+1}(x_r)\right]\theta_k(x_c)\pi(x_c|x_{c-1})\Bigg\}\notag\\
&=& \sum_{a<b}\int\left[\prod_{r=a}^bdx_r\right]\xi_i^a(x_a)\theta_i(x_a)\theta_j(x_b)\pi(x_b|x_{b-1})\notag\\
&&  \left[\prod_{r=a+1}^{b-1}\pi(x_r|x_{r-1})\phi_{i+1}(x_r)\right]
\beta_j^b(x_b,x_c)\theta_k(x_c)dx_c\notag\\
%\beta_{j,k}^b(x_b)\notag\\
\end{eqnarray}
with $\beta_j^b(x,y)$ given by:
\begin{eqnarray}
\beta_j^b (x,y) &:=& \sum_{c=b+1}^{+\infty} \gamma_j^{b,c}(x,y)\\
\gamma_j^{b,c}(x,y) &:=& \int dx_{b+1}^{c-1} \phi_{j+1}(x_{b+1})\pi(x_{b+1}|x)\pi(y|x_{c-1}) \left[\prod_{r=b+2}^{c-1}\pi(x_r|x_{r-1})\phi_{j+1}(x_r)\right]
\end{eqnarray}
Since the underlying Markov chain is time-invariant, $\beta_j^b(x,y)$ does not depend on $b$ and:
\begin{eqnarray}
\langle U_{i,j}U_{j,k}\rangle &=& \int dxdy \omega_i(x)\theta_j(x)\beta_j(x,y)\theta_k(x)
\end{eqnarray}
And more generally:
\begin{eqnarray}
\left\langle\prod_{r=1}^{k}U_{i_{r-1},i_r}\right\rangle &=& \int dx_1^k\left[\prod_{r=2}^k\beta_{i_{r-1}}(x_{r-1},x_r)\theta_{i_r}(x_r)\right]\omega_{i_0}(x_1)\theta_{i_1}(x_1) 
\end{eqnarray}
We therefore have:
\begin{eqnarray}
\frac{\langle U_{ij}U_{jk}\rangle}{\langle U_{ij}\rangle} &=& \frac{\displaystyle{\int dx \omega_i(x)\theta_j(x)\beta_{j,k}(x)}}{\displaystyle{\int dx\omega_i(x)\theta_j(x)}}\label{eq:U-ijU-jkoverU-ij}
\end{eqnarray}
which clearly depends on $i$, since $\omega_i(x)$ cannot, in general, be eliminated from the nmerator and the denominator of Eq.~(\ref{eq:U-ijU-jkoverU-ij}). Therefore mixing the configurations that are in the same $\mathfrak{C}_k$, but have different preparation histories is not allowed in jFFS. 

\section{Summary of Important Notations}

\allowdisplaybreaks
\begin{longtable}{L{3cm}L{0.1cm}L{0.1cm}L{13cm}}
$A$ &&& Starting (meta)stable basin defined by $\{x\in\mathcal{Q}:\lambda(x)< \lambda_A\}$.\\
$B$ &&& Target (meta)stable basin defined by $\{x\in\mathcal{Q}:\lambda(x)\ge\lambda_B\}$.\\
$\mathfrak{C}_k$ &&& $\{x:\lambda_k\le\lambda(x)<\lambda_{k+1}\}$.\\
$\mathcal{E}_A$ &&& Ensemble of trajectories originating in $A$.\\
$f_N$ &&& Number of necessary FFS iterations for $N$ milestones.\\
$L[X]$ &&& The earliest time that a trajectory in $\mathcal{E}_A$ leaves $A$.\\
$N$ &&& Number of FFS milestones between $\lambda_A$ and $\lambda_B$.\\
$N_c$ &&& Number of $\lambda_0$ crossings in conventional FFS.\\
$\mathscr{P}_0(X)$ &&& Probability density of trajectory $X\in\mathcal{E}_A$.\\
$\mathcal{Q}$ &&& Configuration space.\\
$s(x)$ &&& Landing index of $x$, equals $i$ if $x\in\mathfrak{C}_i$, and $-1$ if $\lambda_A\ll\lambda(x)<\lambda_0$.\\
$s_q$ &&& The number of configurations corresponding to a crossing of $\lambda_0$ in Algorithm~\ref{alg:basin-sim} and $\lambda_{k+1}$ in Algorithm~\ref{alg:ffs-it}, with landing index $q$.
\\
$T_B[X]$ &&& The earliest time $X\in\mathcal{E}_A$ returns to $A$ or reaches $B$ after leaving $A$ at $L[X]$.\\
$T_i[X]$ &&& The earliest time $X\in\mathcal{E}_A$ crosses $\lambda_i$ or returns to $A$ after leaving $A$.\\
$U_{i,j}[X]$ &&& Success indicator on whether a trajectory that has crossed into $\mathfrak{C}_i$ at $T_i[X]$ crosses into $\mathcal{C}_j$ at $T_{i+1}[X]$.\\
$W_B[X]$ &&& Success indicator on whether $x_{T_B[X]}\in B$ or not.\\
$\theta_i(x)$ &&& $\theta_{\mathfrak{C}_i}(x)$.\\
$\theta_S(x)$ &&& Indicator function (1 if $x\in S$, 0 otherwise).\\
$\lambda(\cdot):\mathcal{Q}\rightarrow\mathbb{R}$ &&& Order Parameter.\\
$\lambda_m^{\jmath}$ &&& Average amount of jump (or dip) of $\lambda(\cdot)$ between successive configurations of a trajectory. \\
$\hat{\lambda}_k^{\max}$ &&& Maximum value of order parameter for configurations corresponding to crossings of $\lambda_k$ within an FFS iterator.\\
$\xi_i^n(x)$ &&& Probability that $X_n\equiv(x_0,x_1,\cdots,x_{n-1},x)$, a partial trajectory starting in $A$ and ending in $x$: (i) never returns to $A$ after possibly leaving it at some $x_k, (k<n)$, (ii) never crosses $\lambda_i$ before reaching $x$.\\
$\pi(x|y)$ &&& Markov transition probability that $y\in\mathcal{Q}$ is followed by $x\in\mathcal{Q}$.\\
$\Phi_0$ &&& Flux of trajectories crossing $\lambda_0$ after leaving $A$ (Conventional FFS).\\
$\Phi_{A\rightarrow B}$ &&& Flux of trajectories in $\mathcal{E}_A$ that reach $B$ without returning to $A$.\\
$\Phi_{A\rightarrow i}$ &&& Flux of trajectories in $\mathcal{E}_A$ that reach $\lambda_i$ before returning to $A$.\\
$\phi_i(x)$ &&& $\sum_{j=0}^{i}\theta_{j-1}(x)$.\\
$\Psi_{A\rightarrow i}$ &&& Immediate flux of trajectories that immediately jump into $\mathfrak{C}_i$ after crossing $\lambda_0$ for the first time.\\
$\omega_i(x)$ &&& Probability that a trajectory starting in $A$ and ending in $x$ crosses $\lambda_i$ into $\mathfrak{C}_{i}$ at some point in between, and never crosses $\lambda_{i+1}$ or returns to $A$ prior to reaching $x$.
\end{longtable}

\bibliographystyle{apsrev}
\bibliography{References}

\end{document}